\documentclass[10pt,leqno]{amsart}

\usepackage[utf8]{inputenc}
\usepackage[english]{babel}
\usepackage{graphicx}
\usepackage{tabularx}
\usepackage{xcolor}
\usepackage{amsmath,amssymb,amsthm}
\usepackage{csquotes}
\usepackage{hyperref}
\usepackage{paralist}

\usepackage[margin=1in]{geometry}
\hypersetup{
    colorlinks=true,
    linkcolor=black,
    filecolor=black,
    urlcolor=blue
}

\author{Garrett Young \and Mitchell Riley \and Colleen Mitchell}

\title{Fast-Slow Analysis of a Model For the Stimulation of Enzymatic Activity by a Competitive Inhibitor}
\date{} % no date

\begin{document}

\begin{abstract}
Competitive inhibitors can, paradoxically, stimulate an enzymatic reaction at low to moderate
doses. Competitive inhibition of an enzyme occurs when an inhibitor binds to the enzyme’s
binding site and blocks the enzyme’s target molecule from binding. We recently proposed a
detailed but straightforward mass action model for competitive inhibition of phosphoglycerate
kinase 1 (PGK1) by Terazosin (TZ). The full PGK1 model has two substrates and two products which can be bound and released in either order, known as a random bi-bi mechanism. This model, with no further meddling, predicts an increased reaction rate at
low or moderate TZ doses, suggesting that stimulation is an intrinsic feature of competitive inhibition in enzymes with two products. This
mechanism can aid in development of novel therapies, particularly since enzyme activators are
more rare and difficult to design than inhibitors. Here we propose a three time scale reduction of that detailed model and show that the resulting rate equation retains three essential attributes of competitive inhibitor stimulation. These attributes are the biphasic dose response, the dependence on the relative rates of product dissociation from the binary and ternary complexes, and the parameter region where stimulation is possible. The resulting rate equation is a rational function which is a monod function of each substrate, but quadratic in the denominator as a function of inhibitor dose.

\textbf{Relevance to Life Sciences:} TZ, an FDA approved drug for benign prostatic hyperplasia, has neuroprotective properties, likely due to its ability to increase adenosine triphosphate (ATP) production. The original detailed mass action model supported these findings by showing that a competitive inhibitor could stimulate product formation. To clarify this mechanism, we propose a mathematical framework that explains how stimulation arises in this specific competitive inhibition reaction. Such an understanding not only provides a rationale for the observed neuroprotective properties of TZ but also suggests a strategy for identifying other drug–enzyme pairs where enhanced enzyme activation is therapeutically desirable.

\textbf{Mathematical Content:} Differential equations guided by the law of mass action formed the foundation of the original model. Following non-dimensionalization, simulations revealed three distinct timescales. Fast–slow analysis, also known as singular perturbation theory, was used to investigate the fast, slow and super-slow timescales. We derive a single reaction rate equation under conditions consistent with enzyme assay experiments across a range of TZ doses. The reaction rate equation is shown to replicate the behavior of the original model across a range of parameters and doses. 
\end{abstract}

\maketitle
\pagestyle{plain}

\section{Introduction}

Enzymes are proteins that catalyze biochemical reactions. Nearly all reactions in the cell are associated with enzymes and currently over 6,000 enzymes have been classified by their substrates and products and assigned an Enzyme Commission number \cite{McDonald}. These enzymatic reactions can be regulated by physiological factors such as pH, temperature, and substrate concentrations as well as by feedback mechanisms including stimulation and inhibition.  Broadly, an estimated 29\% of all Food and Drug Administration (FDA) approved drugs target enzymes \cite{Santos}. One common type of enzyme regulation is competitive inhibition, which occurs when an inhibitor binds to the enzyme's active site and blocks the enzyme's target molecule from binding.  It is a natural element of the feedback for many enzymes and is the primary mechanism for several classes of medicines, including for example, statins, penicillins and some non-steroidal anti-inflammatory drugs (NSAIDS) \cite{Endo, Yocum, Orlando}. Competitive inhibitors can, paradoxically, stimulate an enzymatic reaction at low to moderate doses \cite{Theorell, Poulikakos, Hatzivassiliou, Hall}. 
We recently proposed a detailed but straightforward mass action model for competitive inhibition of the enzyme phosphoglycerate kinase 1 (PGK1) \cite{Riley}.  This model, with no further meddling, exhibits competitive inhibitor stimulation (CIS) suggesting that CIS is an intrinsic feature of competitive inhibition. The goal of this paper is to understand the mechanism of CIS in the detailed model, and to use that intuition to derive a reduced model which captures the essential features of the full model. 

\subsection{Motivation: Terazosin and Neurodegenerative Disease} 

Neurodegenerative diseases pose an enormous personal and public health challenge. However, we lack treatments to slow or prevent the progressive neuron destruction in Parkinson’s disease (PD), Alzheimer’s disease, Huntington’s disease, and amyotrophic lateral sclerosis. These diseases have the common feature that energy metabolism is impaired \cite{Saxena, Powers, Firbank, Tang}. Previous studies showed that terazosin (TZ), an FDA-approved drug developed to treat hypertension and benign prostatic hyperplasia, has an additional target, phosphoglycerate kinase 1 (PGK1), the first adenosine triphosphate (ATP) generating enzyme in glycolysis \cite{Chen}.

The glycolytic enzyme PGK1 catalyzes the reversible phosphotransfer from 1,3-bisphosphoglycerate (BPG) to adenosine diphosphate (ADP), producing 3-phosphoglycerate (PG) and ATP \cite{Lallemand}. By stimulating PGK1 activity, TZ can enhance energy metabolism in two ways: PGK1 produces ATP itself, and PG serves as a glycolytic substrate leading to increased oxidative phosphorylation and ATP production \cite{Chen, Cai}. Previous work showed that TZ can increase cellular ATP levels in cell culture and animals, and it prevented cell death in PD and other models of neurodegenerative disease \cite{Cai, Boyd, Schultz, Chen2, Chaytow, Weber}. Recently, Kokotos et al suggest that PGK1 is rate limiting in axonal glycolysis of striatal dopaminergic neurons and further show that modest increases in PGK1 activity or PGK1 expression provide a sufficient boost in neuronal ATP kinetics in vivo to protect against axonal dysfunction \cite{Kokotos}. Moreover, multiple investigations of large databases of human patients indicate that the use of TZ is associated with reduced symptoms and delayed onset of PD \cite{Cai, Simmering, Gros, Sasane, Simmering2}.

However, a mechanism by which TZ might activate PGK1 has remained perplexing. The crystal structure of PGK1 and TZ showed that TZ binds to PGK1 at its ADP/ATP binding pocket with the quinazoline portion of TZ overlapping with the adenine ring of ADP \cite{Chen}. That location predicts that TZ would inhibit PGK1 activity by preventing ADP binding, likely as a competitive inhibitor. Consistent with that prediction, high concentrations of TZ inhibited the activity of isolated PGK1 protein and decreased ATP production in cultured cells \cite{Cai}. The relationship between TZ concentrations and activity of PGK1 has also been puzzling; low concentrations of TZ stimulate PGK1 activity, and high concentrations inhibit PGK1 activity. This unusual biphasic activity pattern has been observed in ATP levels not only in isolated enzyme assays but also in both cultured cells and animal models \cite{Chen, Cai}.

\subsection{The Original Detailed Model For PGK1}

In our recent paper, we proposed a detailed model for PGK1 with TZ acting as a competitive inhibitor \cite{Riley}. The model uses mass action kinetics to track the binding of substrates, products and TZ with the enzyme PGK1. Kinetic parameters were estimated from earlier studies and initial conditions were chosen to mirror in vitro enzyme assays \cite{Lallemand, Chen}. For clarity, we will refer to this system and parameters as the original model. Simulations of the original model were used to calculate the initial rates of product formation at quasi-steady-state conditions, which would be established before significant conversion of substrate to product. Surprisingly, and without the need for additional mechanisms, this model qualitatively reproduces the biphasic dose dependence and suggests a novel bypass mechanism for CIS.

\begin{figure}
    \centering
     \includegraphics[width=.8\linewidth]{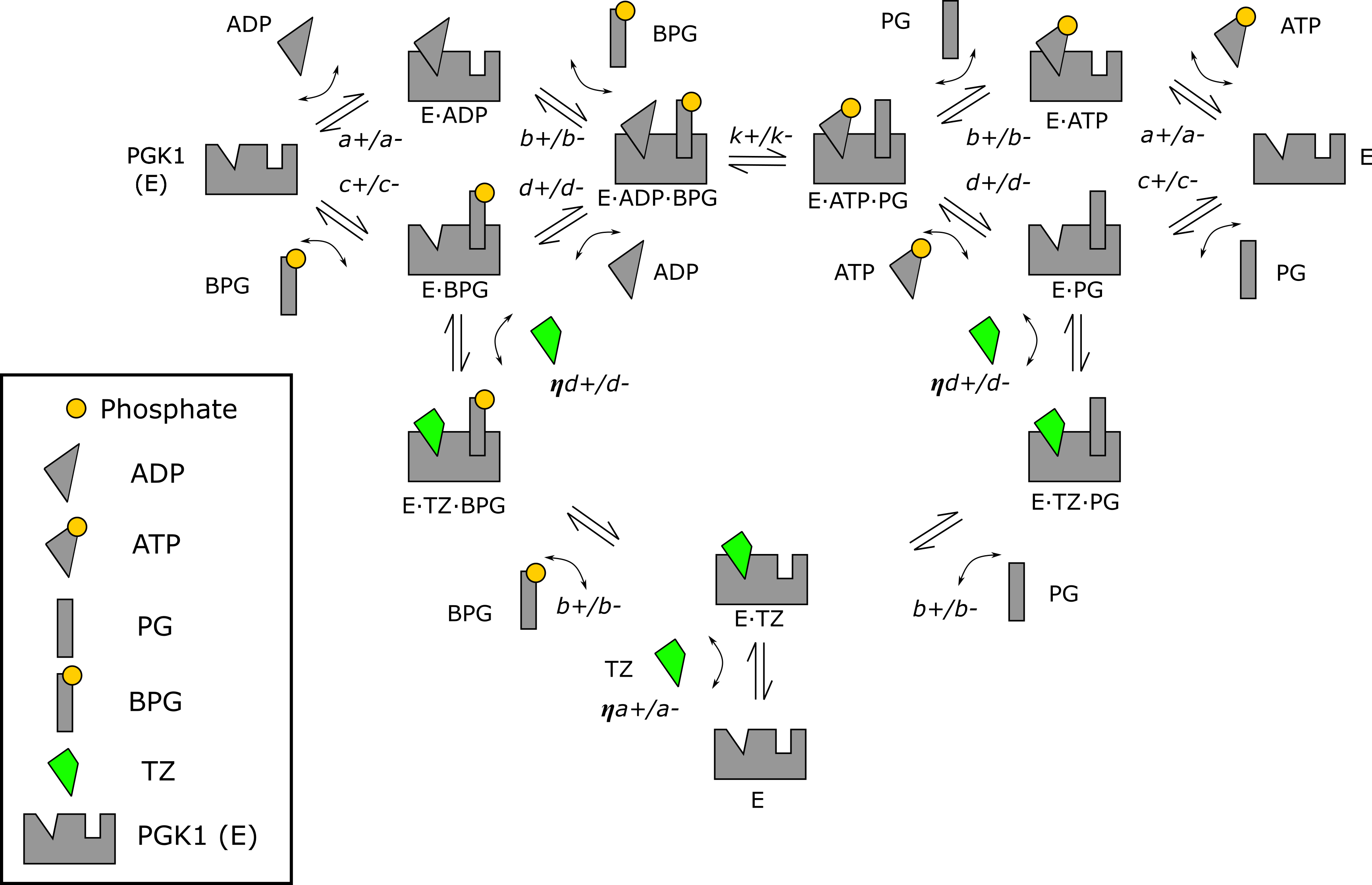}
    \caption{A mass action model describes the interactions of PGK1 with substrates, products, and TZ. The parallel arrows indicate reversible interactions. The Upper Left diamond refers to PGK1’s interaction with substrates ADP and BPG which can bind in either order. The Upper Right diamond refers to PGK1’s interaction with products ATP and PG also in either order. TZ (green) competes with ADP and ATP for their respective binding pocket on PGK1. Figure adapted from \cite{Riley}}
     \label{GeneralModel}
 \end{figure}

The original model tracks changes in concentrations of the unbound enzyme as well as complexes of the enzyme with the substrates, products and competitive inhibitor. For clarity, we refer to PGK1 as E and bound PGK1 as E$\cdot$molecule, e.g., PGK1 bound to ADP is E$\cdot$ADP. Concentrations will be denoted with square brackets, e.g., $[E\cdot ADP]$. Starting at the top left and working left to right in Figure \ref{GeneralModel}, the enzyme (shown with two empty binding sites) can bind the substrates in either order to form complexes E$\cdot$ADP or E$\cdot$BPG before binding to the second substrate to form the ternary complex E$\cdot$ADP$\cdot$BPG. This ternary complex next undergoes the phosphotransfer reaction in which a phosphate from BPG is moved to the ADP, creating a complex with both products E$\cdot$ATP$\cdot$PG. Finally these products dissociate in either order leaving the enzyme available to repeat the reaction steps. The bottom loop of the diagram shows the possible bindings of TZ (green trapezoid) to the ADP/ATP site.

The terms in the original model equations represent the reactions depicted in Figure \ref{GeneralModel}, the rates of which are determined by mass action kinetics. All of the reactions are reversible. The substrates and products are given initial conditions consistent with experiments described in Chen \cite{Chen}. The one notable exception is that the substrate BPG is treated as constant. BPG is unstable in solution. Therefore, a precursor, glyceraldehyde 3-phosphate (GAP) and the enzyme glyceraldehyde 3-phosphate dehydrogenase (GAPDH) are included in the experimental media. The GAPDH enzyme converts one molecule of GAP into one molecule of BPG. This conversion also reduces one molecule of nicotinamide adenine dinucleotide (NAD) from NAD+ to NADH. This is relevant for understanding the model for a few reasons. First, it means that the added concentration of GAP is substantially higher than [BPG]. Second, it means that [BPG] remains essentially constant in the enzyme activity experiment because as it is consumed by the PGK1 reaction, it is replenished by the GAPDH reaction. The value of [BPG] is therefore set to the steady state concentration which would result from the equilibrium of the GAPDH reaction. Third, the total NADH produced by GAPDH, is approximately the same as the amount of ATP produced by PGK1, \cite{Jin}. Finally, because NADH absorbs light strongly at approximately 340 nm, the measurement of absorbance at this wavelength gives a quantification of NADH concentration. Thus, the changes in this absorbance are used to measure the activity of the PGK1 enzyme.

The resulting non-linear ordinary differential equation (ODE) system is shown in equation \ref{RileyModel}. The system has 14 equations corresponding to the free enzyme, the 9 complexes with enzyme bound to other molecules, one of the substrates, ADP, the 2 products, ATP and PG, and the competitive inhibitor,TZ. Since [BPG] is estimated to be constant, there are only three conserved quantities corresponding to the total enzyme, the total ATP and ADP concentration, and the total TZ. The kinetic rate parameters are the bimolecular association and dissociation rate constants determined by Lallemand et al \cite{Lallemand}. The parameters and initial conditions can be found in \cite{Riley} and are also provided in Table \ref{tab:original_params} and \ref{tab:original_initial_conditions} for reference. They correspond to the association/dissociation of ADP and BPG to the various PGK1 forms (E, E$\cdot$ADP, E$\cdot$BPG, and E$\cdot$ADP$\cdot$BPG).

Because the association parameters for ATP and PG have not been experimentally determined, we used the corresponding parameters for ADP and BPG. For example, we set parameter $c \pm$, which describes BPG association or dissociation to E and E$\cdot$BPG, to also describe association or dissociation of PG to E and E$\cdot$PG. The unknown phosphotransfer rate parameters $k_{+}$ and $k_{-}$ are both assigned the value 5.0 $s^{-1}$. Finally, the unitless parameter $\eta$ is TZ’s association parameter factor. Compared to ADP and ATP, TZ has greater binding to PGK1 as indicated by measurements of the dissociation constant \cite{Chen, Flachner}. We will use this original parameter set as an illustration in the simulations and calculations that follow. However, these precise values are not essential. The model can be generalized to a larger class of enzymes with different parameters and the qualitative biphasic behavior is robustly present in a large portion of the parameter space. In Sections 3 and 4, we will explore the impact of key parameters on enzyme activity in both the original model and in the reaction rate equation derived in Section 2. 

\begin{table}[htbp]
  \centering
        \begin{tabularx}{0.75\textwidth} { 
          | >{\raggedright\arraybackslash}X 
          | >{\centering\arraybackslash}X 
          | >{\raggedleft\arraybackslash}X | }
         \hline
         Constants & Forward $(+)$ $(\mu M^{-1}s^{-1})$ & Backward $(-)$  $(s^{-1})$ \\
         \hline
         $a$  & $6.1 \pm 0.3$ & $38 \pm 10$ \\
         \hline
         $b$  & $170 \pm 30$  & $160 \pm 30$ \\
         \hline
         $c$  & $450 \pm 40$  & $14 \pm 26$ \\
         \hline
         $d$  & $4.1 \pm 0.5$ & $270 \pm 30$ \\
         \hline
         $k$  & $ 5.0 (s^{-1})$ & $ 5.0 $ \\
         \hline
         $\eta$  & $ 10^{2.85} $ &  \\
         \hline
         \end{tabularx}
\caption{Reaction rate parameters in the original model. These are the values that were used in \cite{Riley}
 and are used as an example in figures below.}
\label{tab:original_params}
   \end{table}

   \begin{table}[htbp]
  \centering
        \begin{tabularx}{0.75\textwidth} { 
          | >{\raggedright\arraybackslash}X 
          | >{\centering\arraybackslash}X | }
         \hline
         Initial Conditions & Concentrations \\
         \hline
         $[ADP](0)$ & $1000 \; \mu$M \\
         \hline
         $[PGK1](0)$ & $0.04 \; \mu$M \\
         \hline
         $[BPG](0)$ & $80 \; \mu$M \\
         \hline
         $[TZ](0)$ & $2.5$ nM - $25 \; \mu$M \\
         \hline
         \end{tabularx}
    \caption{Initial conditions utilized in the original model. All remaining initial conditions are set to 0. These values are the same as those used in \cite{Riley} and are intended to correspond to the enzyme assay experiments in \cite{Chen}. In particular, they include a small amount of the enzyme, large concentrations of the substrates and a wide range of $[TZ]$ doses.}
    \label{tab:original_initial_conditions}
   \end{table}

\begin{eqnarray}\label{RileyModel}
\nonumber \frac{d[E]}{dt} &=& -a_+ [E] [ADP] - c_+ [E] [BPG] + a_- [E \cdot ADP] + c_- [E \cdot BPG] - a_+ [E] [ATP] \\
\nonumber & & - c_+ [E] [PG] + a_- [E \cdot ATP] + c_- [E \cdot PG] - \eta a_+ [E] [TZ] + a_- [E \cdot TZ] \\
\nonumber \frac{d[ADP]}{dt} &=& -a_+ [E] [ADP] - d_+ [ADP] [E \cdot BPG] + a_- [E \cdot ADP] + d_- [E \cdot ADP \cdot BPG] \\
\nonumber \frac{d[BPG]}{dt} &=& 0 \\
\nonumber \frac{d[E \cdot ADP]}{dt} &=& - a_- [E \cdot ADP] - b_+ [BPG] [E \cdot ADP] + a_+ [E] [ADP] + b_- [E \cdot ADP \cdot BPG] \\
\nonumber \frac{d[E \cdot BPG]}{dt} &=& - c_- [E \cdot BPG] - d_+ [ADP] [E \cdot BPG] + c_+ [E] [BPG] + d_- [E \cdot ADP \cdot BPG] \\
\nonumber & & - \eta d_+ [E \cdot BPG] [TZ] + d_- [E \cdot TZ \cdot BPG] \\
\nonumber \frac{d[E \cdot ADP \cdot BPG]}{dt} &=& -b_- [E \cdot ADP \cdot BPG] - d_- [E \cdot ADP \cdot BPG] + b_+ [BPG] [E \cdot ADP] \\
\nonumber & & + d_+ [ADP] [E \cdot BPG] - k_+ [E \cdot ADP \cdot BPG] + k_- [E \cdot ATP \cdot PG] \\
\nonumber \frac{d[E \cdot ATP \cdot PG]}{dt} &=& - b_- [E \cdot ATP \cdot PG] - d_- [E \cdot ATP \cdot PG] + b_+ [PG] [E \cdot ATP] \\
\nonumber & & + d_+ [ATP] [E \cdot PG] - k_- [E \cdot ATP \cdot PG] + k_+ [E \cdot ADP \cdot BPG] \\
\nonumber \frac{d[PG]}{dt} &=& -b_+ [PG] [E \cdot ATP] - c_+ [E] [PG] + b_- [E \cdot ATP \cdot PG] + c_- [E \cdot PG] \\
 & &  - b_+ [PG] [E \cdot TZ] + b_- [E \cdot TZ \cdot PG] \\
\nonumber \frac{d[ATP]}{dt} &=& -a_+ [E] [ATP] - d_+ [ATP] [E \cdot PG] + a_- [E \cdot ATP] + d_- [E \cdot ATP \cdot PG] \\
\nonumber \frac{d[E \cdot ATP]}{dt} &=& -a_- [E \cdot ATP] - b_+ [PG] [E \cdot ATP] + a_+ [E] [ATP] + b_- [E \cdot ATP \cdot PG] \\
\nonumber \frac{d[E \cdot PG]}{dt} &=& -c_- [E \cdot PG] - d_+ [ATP] [E \cdot PG] + c_+ [E] [PG] + d_- [E \cdot ATP \cdot PG] \\
\nonumber & & - \eta d_+ [E \cdot PG] [TZ] + d_- [E \cdot TZ \cdot PG] \\
\nonumber \frac{d[TZ]}{dt} &=& -\eta d_+ [E \cdot BPG] [TZ] - \eta d_+ [E \cdot PG] [TZ] - \eta a_+ [E] [TZ] + a_- [E \cdot TZ] \\
\nonumber & & + d_- [E \cdot TZ \cdot BPG] + d_- [E \cdot TZ \cdot PG] \\
\nonumber \frac{d[E \cdot TZ]}{dt} &=& -b_+ [BPG] [E \cdot TZ] - b_+ [PG] [E \cdot TZ] - a_- [E \cdot TZ] + b_- [E \cdot TZ \cdot BPG] \\
\nonumber & & + b_- [E \cdot TZ \cdot PG] + \eta a_+ [E] [TZ] \\
\nonumber \frac{d[E \cdot TZ \cdot BPG]}{dt} &=& -d_- [E \cdot TZ \cdot BPG] - b_- [E \cdot TZ \cdot BPG] + \eta d_+ [E \cdot BPG] [TZ] \\
\nonumber & & + b_+ [BPG] [E \cdot TZ] \\
\nonumber \frac{d[E \cdot TZ \cdot PG]}{dt} &=& -d_- [E \cdot TZ \cdot PG] - b_- [E \cdot TZ \cdot PG] + \eta d_+ [E \cdot PG] [TZ] + b_+ [PG] [E \cdot TZ]
\end{eqnarray}

\subsection{Qualitative Behavior and Computational Results}

Surprisingly, no special adaptations of the original model were required to observe an increase in enzyme activity at moderate doses. This model, which describes competitive inhibition in a two substrate enzyme using straightforward mass action, already exhibits the biphasic dose response. In simulations of the original model, the rate of ATP production rises quickly in a matter of seconds and then decays very slowly over several hours as the substrates are depleted and products accumulate. In Chen et al, NADH absorbance was measured at one minute after initiation of the reaction \cite{Chen}. Because we want to focus on the reaction rate in this quasi steady state and for ease of comparison, we computed the average rate of ATP production over the first minute for each simulated dose. We observed a biphasic dose response with an increased ATP production at moderate doses and a decreasing ATP production at high doses.

Further simulations described in \cite{Riley} suggested that the critical feature is the bypass of the rate-limiting release of PG. After phosphotransfer, ATP releases from E$\cdot$ATP$\cdot$PG forming E$\cdot$PG. At a moderate TZ dose, the strong association of TZ forms E$\cdot$TZ$\cdot$PG, and E$\cdot$PG decreases. This changes the configuration of the enzyme for PG release from E$\cdot$PG to E$\cdot$TZ$\cdot$PG and the release parameter for PG from $c_-$ to $b_-$. PG releases more readily from E$\cdot$TZ$\cdot$PG than from E$\cdot$PG because $b_-$ is greater than $c_-$. Then, the clockwise progression of the TZ-bound enzyme around the new cycle ultimately increases E$\cdot$ADP$\cdot$BPG, leading to another round of enzymatic phosphotransfer. This new cycle circumvents the slow release of PG from E$\cdot$PG at the end of the PGK1 reaction.
In contrast, at the high dose, TZ acts as a true competitive inhibitor. With TZ binding most of the PGK1,
there is very little non-TZ-bound enzyme to allow the reaction to take place. These computational results hint at the importance of the relationship between the dissociation rate of PG from E$\cdot$PG, $c_-$, and from E$\cdot$TZ$\cdot$PG, $b_-$. Simulations of the full model with varying $c_-$, but fixing $b_-$, showed that when the $c_-/b_-$ ratio is close to zero, all TZ concentrations stimulated ATP production. However, as the $c_-/b_-$ ratio increases, stimulation ceases at moderate TZ concentrations. 

The goal of this paper is to use this intuition to reduce the original model to a simple reaction rate equation which captures the essential features and provides insight into the mechanism of CIS in general. The methods and model details will be split into Sections 2, 3, and 4. Section 2 details the steps of the reduction. In section 2.1 we non-dimensionalize the system. Next, in 2.2, we identify three timescales and define two small parameters. In Section 2.3 we complete the reduction in four steps. First, we eliminate any edges in the reaction diagram which are not essential to the mechanism. Next we focus on the middle, slow, time scale by making both equilibrium and quasi-steady state approximations. Finally, we return to the super slow time scale to write the reaction rate as a rational function of substrates, TZ dose and parameters. 

In section 3, we examine the qualitative behavior of the reaction rate equation. We derive a region of parameter space which allows stimulation. Next, we investigate two doses with potential clinical importance, the optimal dose and the maximum stimulating dose. We conclude our investigate of the behavior of the reaction rate equation by comparing to classic reaction rate models. 

Finally, in section 4 we compare numerical solutions of the dimensionless model to the reaction rate equation. In section 4.1 we show that the reaction rate equation retains the same region of parameter space where the biphasic dose response is observed. In section 4.2, we return to the dependence on the $c_-/b_-$ ratio and show that the reaction rate equation retains the qualitative features of the original model.

\section{Asymptotic Reduction as a Three Time Scale System}

While we have gained intuition from exploration of the original model, it can be unwieldy. The goal of this section is to approximate the original model with a reduced model in the form of a reaction rate equation that can be understood analytically. The first step in section 2.1 will be to non-dimensionalize the system. Next in section 2.2, we identify the three timescales inherent in the dynamics and use the separation in these timescales to define two small parameters, $\varepsilon_1$ and $\varepsilon_2$. Finally, in section 2.3 we will complete the reduction in four steps. The first step is to argue that the random bi-bi mechanism (where the substrates and products can bind and release in either order) can be replaced with an ordered mechanism in which BPG binds before ADP and ATP is released before PG. Next we consider the small $\varepsilon_1$ limit and treat the super-slow variables as constants. The small $\varepsilon_2$ limit gives a system of equilibrium approximations for the dynamics in the fast scale. Reparameterizing on the slow manifold gives a single ODE. Finally, we investigate the quasi-steady state approximation for this equation and derive an equation for the reaction rate. This reaction rate equation retains the qualitative behavior of the full model and elucidates the relevant relationships among parameters.

\subsection{Non-dimensionalization}

\begin{figure}
 %\begin{wrapfigure}{R}{0.4\textwidth}
 %\vspace{-20pt}
     \centering
     \includegraphics[width=.8\linewidth]{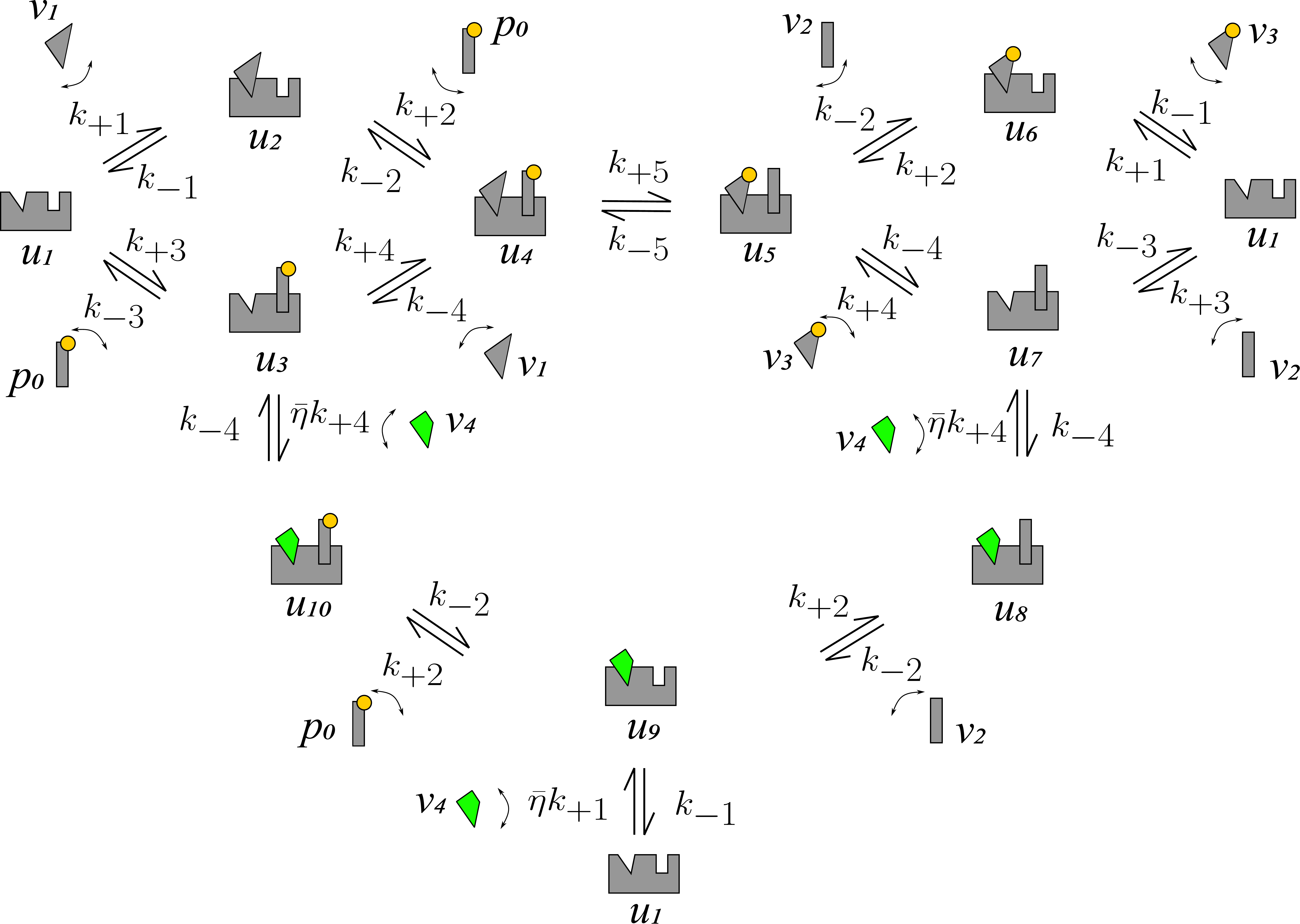}
     \caption{The dimensionless mass action model derived from Figure \ref{GeneralModel}. Proportions of enzyme and enzyme complexes are denoted with $u_1, \dots, u_{10}$ and the proportions of unbound substrates and products are denoted $v_1, \dots v_4$. The rate constants shown here are for the rates of change of enzyme complexes $u_1, \dots, u_{10}$. Due to different scalings for the unbound species, the differential equations for the variables $v_1, \dots v_4$, must be scaled appropriately.}
      \label{Non-DimModel}
%  %\end{wrapfigure}
  \end{figure}

To begin our rescaling, we observe that there are three conserved quantities, the total enzyme, the total ATP \& ADP and the total TZ. The total enzyme, unbound and in complex is $e_0$,
\begin{align*}
    e_0 &= [E] + [E \cdot ADP] + [E \cdot BPG] + [E \cdot ADP \cdot BPG] + [E \cdot ATP \cdot PG] + [E \cdot ATP] + [E \cdot PG] + \dots \\
    & \hspace{1.5in} [E \cdot TZ \cdot PG] + [E \cdot TZ] +  [E \cdot TZ \cdot BPG]
\end{align*}
Each of these concentrations is scaled with $e_0$ to give the proportion of total enzyme. These proportions are denoted with $u_1, \cdots, u_{10}$. Thus, we have $u_1=[E]/e_0$, $u_2=[E \cdot ADP]/e_0$, $u_3 = [E \cdot BPG]/e_0$, $u_4 = [E \cdot ADP \cdot BPG]/e_0$, $u_5 = [E \cdot ATP \cdot PG]/e_0$, $u_6 = [E \cdot ATP]/e_0$, $u_7 = [E \cdot PG]/e_0$, $u_8 = [E \cdot TZ \cdot BPG]/e_0$, $u_9 = [E \cdot TZ]/e_0$, and $u_{10} = [E \cdot TZ \cdot BPG]/e_0$. For ease of reference, we include in Figure \ref{Non-DimModel} a version of the reaction network diagram labeled with the dimensionless variables.  

It remains to find reasonable scalings for the concentrations of the chemical species not bound to the enzyme. One of the other conserved quantities is the total of the bound and unbound ATP and ADP, $a_0$.
\begin{equation*}
a_0 = [ADP] + [ATP] + [E \cdot ADP] + [E \cdot ATP] + [E \cdot ADP \cdot BPG] + [E \cdot ATP \cdot PG].
\end{equation*}
We therefore define the scaled ADP as $v_1=[ADP]/a_0$ and scaled ATP as $v_3=[ATP]/a_0$. $v_1$ and $v_3$ give the proportion of the total ADP and ATP that is not bound to the enzyme. In contrast, the total BPG and PG is not conserved because BPG is replenished as it is consumed so $[BPG]$ is constant. Therefore, we set $p_0=[BPG]$ and scale PG in comparison, $v_2=[PG]/p_0$. Finally, the total bound and unbound TZ is conserved.
\begin{equation*}
z_0 = [TZ] + [E \cdot TZ \cdot PG] + [E \cdot TZ] +  [E \cdot TZ \cdot BPG]
\end{equation*}
We set $v_4=[TZ]/z_0$ to be the proportion of total TZ which is unbound. 

Time is scaled to the central phosphotransfer step in the reaction, $\xi = k_+ t$. We can think of this as the key step which creates new ATP. This step defines what we will call the slow time scale in the next section. 

The final step of rewriting the system is to simplify the notation for the reaction rates. We number the reactions and denote association reaction rates with $+$ and dissociation reaction rates with $-$, see Figure \ref{Non-DimModel}. The dimensionless rate parameters are given in Table \ref{tab:new_params}. 
%k_{+3} = \frac{c_+ p_0}{k_+}$, $k_{-3} = \frac{c_-}{k_+}$, $k_{+4} = \frac{d_+ a_0}{k_+}$, $k_{-4} = \frac{d_-}{k_+}$, and $\bar{\eta} = \frac{\eta z_0}{a_0}$.
In the original model, the association of TZ is stronger than that for ADP or ATP, necessitating the additional $\bar{\eta}$ parameter in the reactions which bind TZ. %As an example, the equation for the rescaled $[E \cdot BPG]$ variable is 
%$$\dot{u}_3 = k_{+3} u_1 - k_{-3}u_3 - k_{+4} u_3 v_1 + k_{-4} u_4 - \bar{\eta} k_{+4} u_3 v_4 + k_{-4}u_{10}$$ 
Note that equations for the species not bound to enzyme ($v_1, v_2, v_3$, and $v_4$), the rates are scaled appropriately. %For example 
%$$\dot{v}_1 = \frac{e_0}{a_0}(- k_{+1} u_1 v_1 + k_{-1} u_2 -  k_{+4} u_3 v_1 + k_{-4} u_4).$$
The full set of equations in the dimensionless system are given in equation \ref{dimensionless}.

\begin{table}[htbp]
  \centering
        \begin{tabularx}{0.75\textwidth} { 
          | >{\raggedright\arraybackslash}X 
          | >{\centering\arraybackslash}X 
          | >{\centering\arraybackslash}X | }
          \hline
         Constants & Forward $(+)$ & Backward $(-)$  \\
         \hline
         $k_{\pm 1}$ & $\frac{a_+ a_0}{k_+} = 1220$ & $\frac{a_-}{k_+} = 7.6$ \\
         \hline
         $k_{\pm 2}$ & $\frac{b_+ p_0}{k_+} = 2720$ & $\frac{b_-}{k_+} = 32$ \\
         \hline
         $k_{\pm 3}$ & $\frac{c_+ p_0}{k_+} = 7200$ & $\frac{c_-}{k_+} = 2.8$ \\
         \hline
         $k_{\pm 4}$ & $\frac{d_+ a_0}{k_+} = 820$ & $\frac{d_-}{k_+} = 54$ \\
         \hline
         $k_{\pm 5}$ & $\frac{k_+}{k_+} = 1$ & $\frac{k_-}{k_+} = 1$ \\
         \hline
         $\bar{\eta}$ & $\frac{\eta z_0}{a_0} =$  \footnotesize{$0.00177$} - \footnotesize{$17.7$} & \\
         \hline
    \end{tabularx}
    \caption{Parameters of the dimensionless model written in terms of the parameters of the original model. The unitless numerical values for the original parameter set are included.}
    \label{tab:new_params}
\end{table}

\begin{align}\label{dimensionless}
\nonumber \frac{u_1}{d\xi} &= -k_{+1} u_1 v_1 - k_{+3} u_1 + k_{-1} u_2 + k_{-3} u_3 - k_{+1} u_1 v_3 - k_{+3} u_1 v_2 + k_{-1} u_6 \\
       \nonumber  &\quad + k_{-3} u_7 - \bar{\eta} k_{+1} u_1 v_4 + k_{-1} u_9 \\
\nonumber \frac{u_2}{d\xi} &= -k_{-1} u_2 - k_{+2} u_2 + k_{+1} u_1 v_1 + k_{-2} u_4 \\
\nonumber \frac{u_3}{d\xi} &= -k_{-3} u_3 - k_{+4} u_3 v_1 + k_{+3} u_1 + k_{-4} u_4 - \bar{\eta} k_{+4} u_3 v_4 + k_{-4} u_{10} \\
\nonumber \frac{u_4}{d\xi} &= -k_{-2} u_4 - k_{-4} u_4 + k_{+2} u_2 + k_{+4} u_3 v_1 - k_{+5} u_4 + k_{-5} u_5 \\
\nonumber \frac{u_5}{d\xi} &= -k_{-2} u_5 - k_{-4} u_5 + k_{+2} u_6 v_2 + k_{+4} u_7 v_3 - k_{-5} u_5 + k_{+5} u_4 \\
\nonumber \frac{u_6}{d\xi} &= -k_{-1} u_6 - k_{+2} u_6 v_2 + k_{+1} u_1 v_3 + k_{-2} u_5 \\
\nonumber \frac{u_7}{d\xi} &= -k_{-3} u_7 - k_{+4} u_7 v_3 + k_{+3} u_1 v_2 + k_{-4} u_5 - \bar{\eta} k_{+4} u_7 v_4 + k_{-4} u_8 \\
\frac{u_8}{d\xi} &= -k_{-4} u_8 - k_{-2} u_8 + \bar{\eta} k_{+4} u_7 v_4 + k_{+2} u_9 v_2 \\
\nonumber \frac{u_9}{d\xi} &= -k_{+2} u_9 v_2 - k_{+2} u_9 + k_{-2} u_8 + k_{-2} u_{10} - k_{-1} u_9 + \bar{\eta} k_{+1} u_1 v_4 \\
\nonumber \frac{u_{10}}{d\xi} &= -k_{-2} u_{10} - k_{-4} u_{10} + k_{+2} u_9 + \bar{\eta} k_{+4} u_3 v_4 \\
\nonumber \frac{v_1}{d\xi} &= \frac{e_0}{a_0} (- k_{+1} u_1 v_1 - k_{+4} u_3 v_1 + k_{-1} u_2 + k_{-4} u_4 ) \\
\nonumber \frac{v_2}{d\xi} &= \frac{e_0}{p_0} (- k_{+2} u_6 v_2 - k_{+3} u_1 v_2 + k_{-2} u_5 + k_{-3} u_7 - k_{+2} u_9 v_2 + k_{-2} u_8 ) \\
\nonumber \frac{v_3}{d\xi} &= \frac{e_0}{a_0} (- k_{+1} u_1 v_3 - k_{+4} u_7 v_3 + k_{-1} u_6 +  k_{-4} u_5) \\
\nonumber \frac{v_4}{d\xi} &= \frac{e_0}{z_0} (- \bar{\eta} k_{+4} u_3 v_4 - \bar{\eta} k_{+4} u_7 v_4 - \bar{\eta} k_{+1} u_1 v_1 + k_{-4} u_{10} + k_{-4} u_8 +  k_{-1} u_9)
\end{align}

\subsection{The Three Timescales}
\begin{figure}
    %\centering
    \includegraphics[width=1\linewidth]{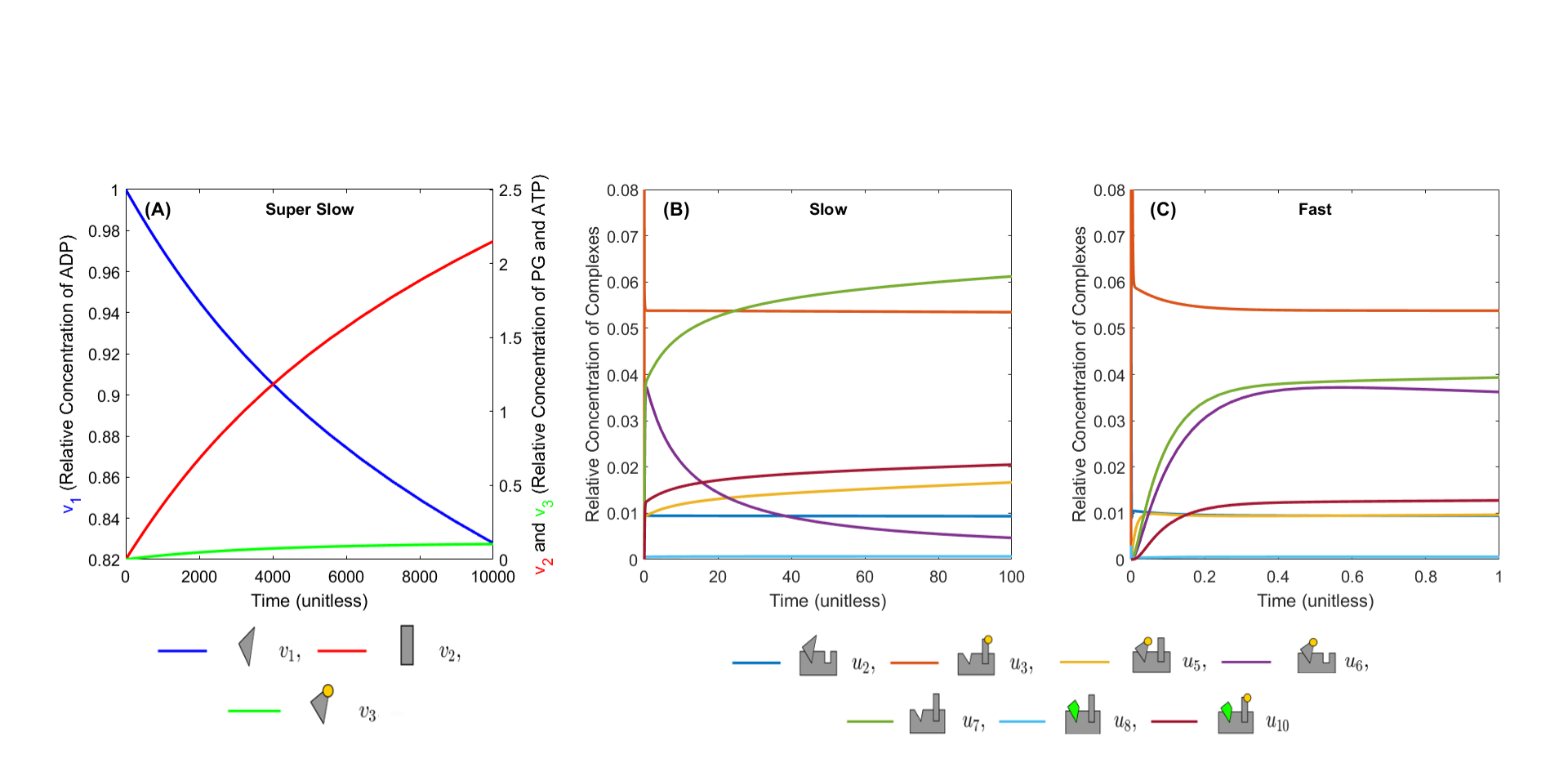}
    \vspace{-15pt}
    
   \caption{Simulations of the dimensionless model reveal three distinct timescales. This figure was made with a moderate drug dose, $z_0 = 50$nM. Panel A shows the super-slow changes to substrate and product concentrations over 10,000 units of time. Note PG and ATP are being produced at the very nearly the same rate, but they were scaled differently since $v_2=[PG]/p_0$ and $v_3=[ATP]/a_0$. The left side of panel A is the scale for $v_1$, while the right is $v_2$ and $v_3$. Panel B shows the changes in the enzyme complexes over 100 time steps. Panel C shows the fastest changes over the first unit of time. Recall that each unit of time is $\frac{1}{k_+}=0.2$ seconds.}
   \label{fig:threescales}
\end{figure}

Based on our exploration of the dimensionless model (equation \ref{dimensionless}), we observe that the dynamics occur on three distinct timescales. Figure \ref{fig:threescales} shows numerical solutions with $z_0 = 50$nM in 10,000, 100 and 1 unit of time. Two additional examples with zero dose, $z_0 = 0$nM, and a high dose, $z_0 = 25 \mu$M are given in the supplement. $v_1, v_2,$ and $v_3$ evolve on a super-slow scale (Panel A).  In contrast, the evolution of the enzyme complexes can be viewed as occurring in two phases, a slow and fast timescale. In Panel C, the fast dynamics approach constant values in a fraction of one unit of time. In Panel B on the slow time scale, the variables approach new slow quasi-steady states over hundreds of time steps. 

The depletion of substrate, $v_1$, and accumulation of products, $v_2$ and $v_3$, take place on the super-slow timescale. This is because the substrates ADP and BPG are more abundant than the total enzyme. In our non-dimensionalization each of these equations includes a scaling factor proportional to the ratio of $e_0$ to $a_0$. 
We capture the dynamics on this super-slow scale with the small parameter $\varepsilon_1 = \frac{e_0}{a_0} k_{-4}$ or $\varepsilon_1=\frac{d_- e_0}{k_+ a_0} \approx 0.00216$ in the parameters of the original model.

The phosphotransfer reaction, $u_4 \rightarrow u_5$, (see Figure \ref{Non-DimModel}) is the mandatory central step which adds the phosphate group to $v_1$ to form $v_3$. The numerical investigation of the original model suggests that the mechanism by which the inhibitor, $v_4$, can accelerate the overall reaction rate is to create a bypass so that the enzyme can follow a faster route to reset after this reaction. When no $v_4$ was present the enzyme flux was strongest along the path $u_5 \rightarrow u_7 \rightarrow u_1 \rightarrow u_3 \rightarrow u_4$. Note in this pathway, when $v_2$ is released from $u_7$, $u_1$ returns to the lefthand side of the diagram in Figure \ref{Non-DimModel}. Along this path we observe that the rate limiting step is the release of $v_2$, $u_7 \rightarrow u_1$. In contrast, at a stimulating dose of $v_4$, the path  with the greatest flux was $u_5 \rightarrow u_7 \rightarrow u_8 \rightarrow u_9 \rightarrow u_{10} \rightarrow u_3 \rightarrow u_4$. Along this path the rate limiting step is also release of $v_2$, $u_8 \rightarrow u_9$. We will therefore think of these three reactions  $u_4 \rightarrow u_5$, $u_7 \rightarrow u_1$, and $u_8 \rightarrow u_9$ as belonging to the slow timescale. We define $\varepsilon_2= \frac{1}{k_{+4}}$, or $ \frac{k_+}{d_+ a_0} \approx 0.00122$ in the parameters of the original model, as the small parameter reflecting that the phosphostransfer step is slower than the substrate binding steps. 

This yields a system that can be written in the form 
\begin{eqnarray*}
\frac{d\mathbf{u}}{d\xi}&=&f(\mathbf{u},\mathbf{v})+\frac{1}{\varepsilon_2}g(\mathbf{u},\mathbf{v})\\
\frac{d\mathbf{v}}{d \xi}&=&\varepsilon_1 h(\mathbf{u},\mathbf{v})
\end{eqnarray*}
where $\mathbf{u}=\{u_1, u_2, \dots,u_{10}\}$ and $\mathbf{v}=\{v_1,v_2,v_3,v_4\}$.
The function $f$ includes the terms for the three reactions on the slow time scale: (1) the phosphotransfer, $u_4 \rightarrow u_5$, (2) the release of $v_2$, $u_7 \rightarrow u_1$ and (3) the faster release of $v_2$ when bound to $v_4$, $u_8 \rightarrow u_9$. The remaining reactions among enzyme complexes are on the fast scale and make up the function $g$. The function $h$ includes the terms for the binding and dissociation of substrates and products on the super-slow time scale. 

It is also important to note that the original model focused on the concentrations one minute into the reaction for comparisons to in vitro experiments. One minute is 300 time units in this scaling and is therefore best approximated with the quasi-steady state of the slow timescale. This is consistent with our choice of timescale, $\xi$, to define the middle or slow timescale.

\subsection{Model reduction}

In this section we will reduce the model in four steps. First we will constrain the substrate binding and product release to occur in a specific order, and remove the reaction $u_9 \leftrightarrow u_1$. Next, in section 2.3.2 we take the limit $\varepsilon_1 \rightarrow 0$ and approximate the super-slow variables as constant. In section 2.3.3 we take the limit $\varepsilon_2 \rightarrow 0$, introducing a collection of equilibrium approximations. This leads to a natural grouping of the variables to create a new parameterization of the slow manifold in two variables. This ODE system of two variables has a conserved quantity allowing the full reduction to one deceptively simple ODE. Finally, in 2.3.4 the quasi-steady state of the slow time dynamics is used to derive an approximation of the rate of the overall reaction. 

\subsubsection{Simplify Random Bi-Bi Mechanism to Ordered Ternary Mechanism}

\begin{figure}
    \centering
    \includegraphics[width=0.9\linewidth]{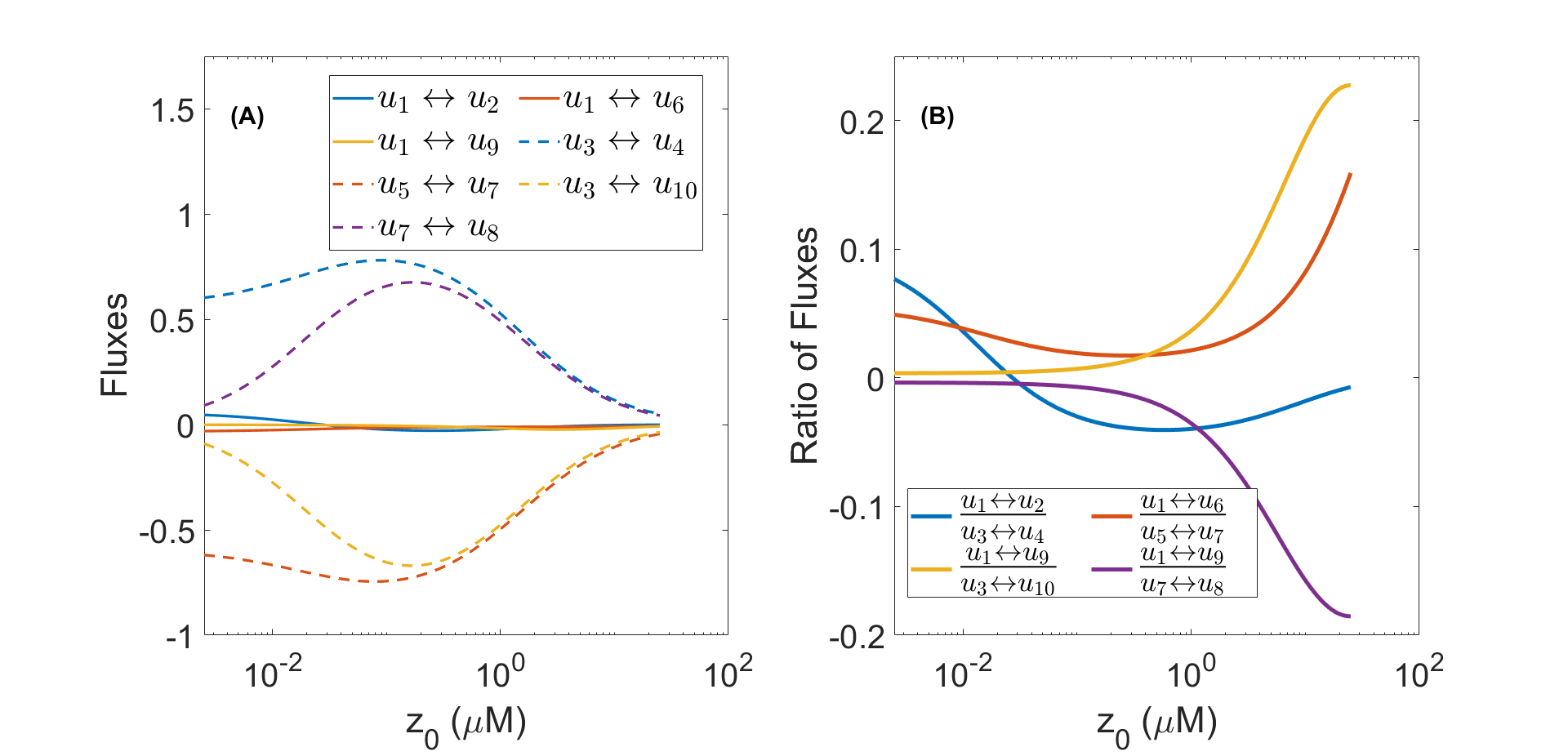}
    \vspace{-5pt}
    
   \caption{Comparison of net fluxes for individual reactions as a function of $[TZ]$. Net flux is defined to be positive in the direction of binding and negative in the direction of dissociation. The fluxes used in both panels are measured at $\xi=300$ dimensionless time units, the equivalent of one minute in the original model. Panel A shows the fluxes of those reactions we will remove and a similar retained reaction. Solid lines represent the net flux of a reaction where a substance is binding to or dissociating to form empty enzyme, $u_1$. The dashed lines represent the corresponding reactions to a non-empty enzyme. The blue curves are the net fluxes of the reactions reversibly binding ADP, $v_1$. The solid blue is ADP binding with E. The dashed blue is ADP with E$\cdot$BPG. The ratio of these net fluxes is the blue curve in panel B and indicates a preference for $u_1 \rightarrow u_3 \rightarrow u_4$ over $u_1 \rightarrow u_2 \rightarrow u_4$. Similarly, the red curves are the reactions binding ATP, $v_3$. The solid red is ATP with E. The dashed red is ATP with E$\cdot$PG. The red curve in panel B is the ratio and indicates a preference for $u_5 \rightarrow u_7 \rightarrow u_1$ over $u_5 \rightarrow u_6 \rightarrow u_1$. The yellow and purple curves are the reactions binding TZ, $v_4$. The solid yellow is TZ binding with E. The dashed yellow is TZ with E$\cdot$BPG. The dashed purple is TZ with E$\cdot$PG. The ratios in yellow and purple in panel B indicate a preference for $u_8 \rightarrow u_9$ and $u_9\rightarrow u_{10}$ over $u_9 \rightarrow u_1$. The ratios of fluxes in panel B are small for low and moderate doses, and still reasonably small for large doses. We observe that the ratios grow for large TZ doses primarily due to a reduction in the denominators. }
   \label{fig:FluxComparison}
\end{figure}

We begin our reduction by simplifying the reaction network by removing those paths in the network that do not have substantial flux and whose elimination does not change the qualitative behavior of the full system. We will eliminate the nodes for $u_2$ and $u_6$ as well as the edge connecting $u_9$ to $u_1$. The removal of nodes $u_2$ and $u_6$ means that we are replacing the original random bi-bi mechanism with an ordered mechanism. In the absence of the inhibitor, $v_4$, the reaction will proceed $u_1 \rightarrow u_ 3 \rightarrow u_4 \rightarrow u_5 \rightarrow u_7 \rightarrow u_1$. In the context of the enzyme PGK1, this means that the enzyme will bind first BPG and then ADP and that the products will dissociate first ATP and then PG. 

\textbf{Removal of $u_2$} 

\begin{figure}
    \centering
    \includegraphics[width=0.8\linewidth]{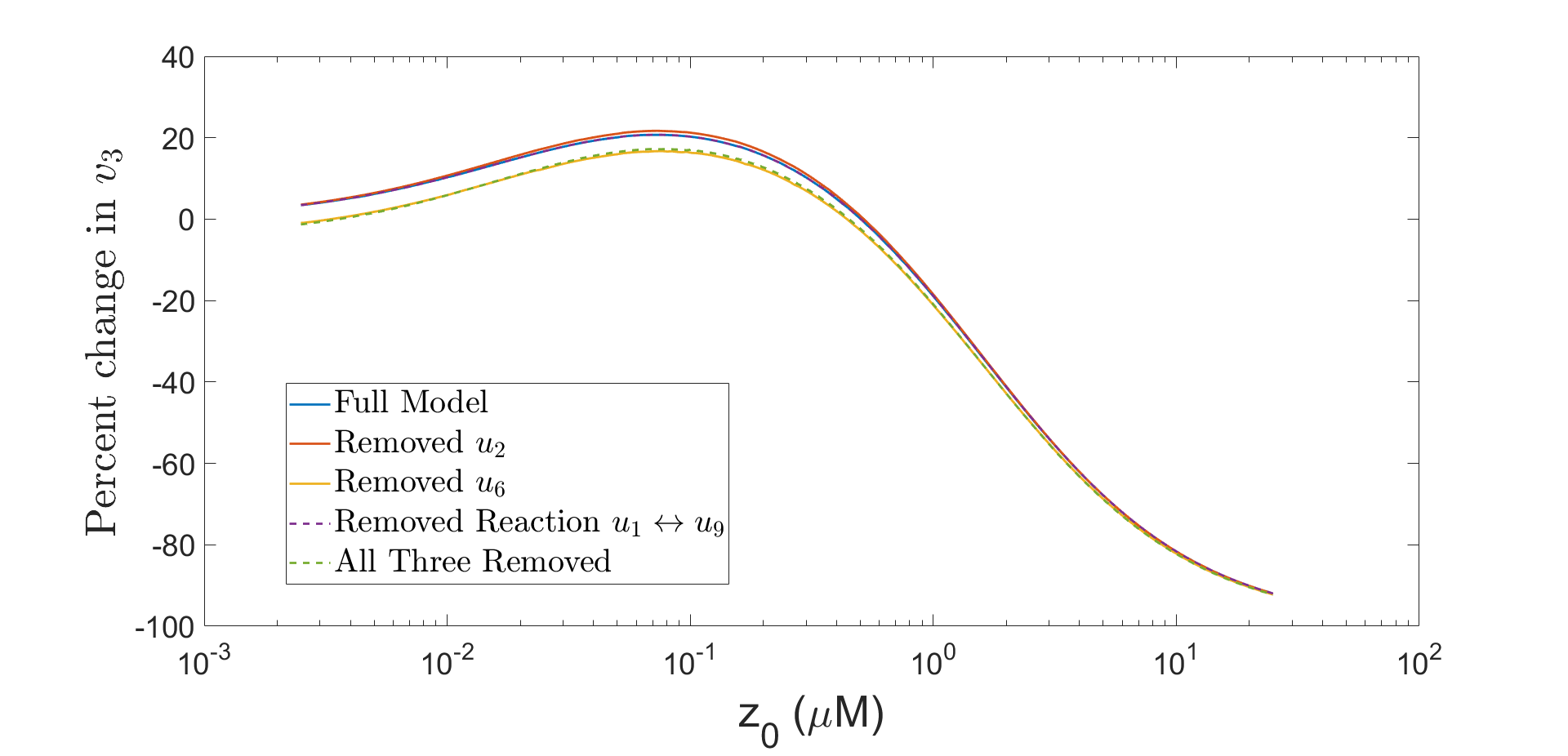}
    \vspace{-5pt}
    
   \caption{The full dimensionless model (shown in blue) already displays stimulation. Upon removal of the specified nodes from the model, the the stimulation of ATP production persists. The removal of all three reaction nodes does not qualitatively change the production of $v_3$.}
   \label{fig:ATPProductionModels}
\end{figure}

Simulations with the dimensionless model suggest that for all doses tested, the fluxes $u_1 \rightarrow u_3$ and $u_3 \rightarrow u_4$ were substantially bigger than the flux from $u_1 \rightarrow u_2 \rightarrow u_4$. The blue curves in both panels of Figure \ref{fig:FluxComparison} compare the net fluxes of the two reactions for binding $v_1$, $u_1 \rightarrow u_2$ and $u_3 \rightarrow u_4$. We see in Panel B that the ratio is always relatively small, indicating that $u_3 \rightarrow u_4$ is more prominent. In our simulations for a wide range of doses, the binding of $p_0$ then $v_1$ is preferred over binding in the opposite order. This is consistent with the large differences in rate constants for the reactions $u_1 \rightarrow u_2$ and $u_1 \rightarrow u_3$. The ratio $k_{+3}/k_{-3}$ is approximately 16 times larger than the ratio $k_{+1}/k_{-1}$. Additionally, in Figure \ref{fig:ATPProductionModels} the removal of the node for $u_2$ does not qualitatively change the dynamics of the system. 

\textbf{Removal of $u_6$} 

For all doses, the fluxes $u_5 \rightarrow u_7$ and $u_7 \rightarrow u_1$ were substantially larger than the flux from $u_5 \rightarrow u_6 \rightarrow u_1$. The red curves in both panels of Figure \ref{fig:FluxComparison} compare the net fluxes of the reactions for the dissociation of $v_2$, $u_5 \rightarrow u_6$ and $u_7 \rightarrow u_1$. The ratio in panel B stays relatively small as the drug dose varies. Therefore, the preferred path is to first release $v_3$ then $v_2$. Again this is consistent with the differences in the reaction rates. The ratio $k_{-4}/k_{+4}$ is approximately 5.6 times larger than the ratio $k_{-2}/k_{+2}$ and because of the different scalings $v_2$ is 12.5 times larger than $v_3$. This alteration to the model does not qualitatively change the biphasic behavior in Figure \ref{fig:ATPProductionModels} but it does slightly decrease the $v_3$ production rate at low and moderate doses. This decrease is noticeable at the $\xi=300$ time point but is even closer to the full model at later times as the super slow accumulation of $v_2$ reduces the net flux from $u_5 \rightarrow u_6$. 

\textbf{Removal of $u_9 \leftrightarrow u_1$} 

There are three reactions involving $v_4$, so we compared the reaction of $u_1 \leftrightarrow u_9$ in Figure \ref{fig:FluxComparison} with $u_3 \leftrightarrow u_{10}$ and $u_7 \leftrightarrow u_8$ in yellow and purple, respectively. No matter the $v_4$ dose, the flux between $u_1$ and $u_9$ was smaller than the other two reactions. Again returning to underlying parameters, the ratio $k_{_2}/k_{-2}$ is approximately 6.8 $\times 10^6$ times larger than the ratio $k_{-1}/\eta k_{+1}$ so even across a large range of $v_4$ doses, the $u_9 \leftrightarrow u_{10}$ reaction will dominate $u_9 \leftrightarrow u_1$. This alteration to the model also does not qualitatively change the behavior of in Figure \ref{fig:ATPProductionModels}.

These three alterations simplify the calculations and retain the qualitative biphasic behavior of the full dimensionless model. The blue line in Figure \ref{fig:ATPProductionModels} shows the percent change in $v_3$ compared to the reaction rate with no drug ($z_0=0$) at $\xi=300$ time units without any alterations. The green dashed line represents removing all three reactions. In all cases, the qualitative behavior is maintained. Small doses slightly increase $v_3$ production, moderate doses lead to a substantial increase, and large doses decrease production. We have plotted here the percent change in $v_3$ in the dimensionless model (equation \ref{Non-DimModel}) to be consistent with the simulations in \cite{Riley} and the enyzyme assay experiments in \cite{Chen}. It is important to note that the rate of production of $v_3$ is approximately constant on this timescale. Therefore, the percent change in $v_3$ can be interpreted as a percent change in the rate of the reaction. The system resulting from these changes is
\begin{align}
    \nonumber \frac{u_1}{d\xi} &= - k_{+3} u_1 v_2 + k_{-3} u_7 + \frac{1}{\varepsilon_2} \left(- \frac{k_{+3}}{k_{+4}} u_1 + \frac{k_{-3}}{k_{+4}} u_3 \right) \\
    \nonumber \frac{u_3}{d\xi} &= \frac{1}{\varepsilon_2} \left(- \frac{k_{-3}}{k_{+4}} u_3 - u_3 v_1 + \frac{k_{+3}}{k_{+4}} u_1 + \frac{k_{-4}}{k_{+4}} u_4 - \bar{\eta} u_3 v_4 + \frac{k_{-4}}{k_{+4}} u_{10} \right) \\
    \nonumber \frac{u_4}{d\xi} &= - k_{+5} u_4 + k_{-5} u_5 + \frac{1}{\varepsilon_2} \left(- \frac{k_{-4}}{k_{+4}} u_4 + u_3 v_1 \right)\\
    \nonumber \frac{u_5}{d\xi} &= - k_{-5} u_5 + k_{+5} u_4 + \frac{1}{\varepsilon_2} \left(- \frac{k_{-4}}{k_{+4}} u_5 + u_7 v_3 \right)\\
    \nonumber \frac{u_7}{d\xi} &= - k_{-3} u_7 + k_{+3} u_1 v_2 + \frac{1}{\varepsilon_2} \left( - u_7 v_3  + \frac{k_{-4}}{k_{+4}} u_5 - \bar{\eta} u_7 v_4 + \frac{k_{-4}}{k_{+4}} u_8 \right) \\
    \nonumber \frac{u_8}{d\xi} &= - k_{-2} u_8 + k_{+2} u_9 v_2 + \frac{1}{\varepsilon_2} \left(- \frac{k_{-4}}{k_{+4}} u_8 + \bar{\eta} u_7 v_4 \right) \\
    \frac{u_9}{d\xi} &= - k_{+2} u_9 v_2 + k_{-2} u_8 + \frac{1}{\varepsilon_2} \left( - \frac{k_{+2}}{k_{+4}} u_9 + \frac{k_{-2}}{k_{+4}} u_{10} \right) \\
    \nonumber \frac{u_{10}}{d\xi} &= \frac{1}{\varepsilon_2} \left(- \frac{k_{-2}}{k_{+4}} u_{10} - \frac{k_{-4}}{k_{+4}} u_{10} + \frac{k_{+2}}{k_{+4}} u_9 + \bar{\eta} u_3 v_4 \right) \\
    \nonumber \frac{v_1}{d\xi} &= \varepsilon_1 \left( - \frac{k_{+4}}{k_{-4}} u_3 v_1 + u_4 \right) \\
    \nonumber \frac{v_2}{d\xi} &= \varepsilon_1 \frac{a_0}{p_0} \left( - \frac{k_{+3}}{k_{-4}} u_1 v_2 + \frac{k_{-3}}{k_{-4}} u_7 - \frac{k_{+2}}{k_{-4}} u_9 v_2 - \frac{k_{-2}}{k_{-4}} u_8 \right) \\
    \nonumber \frac{v_3}{d\xi} &= \varepsilon_1 \left(- \frac{k_{+4}}{k_{-4}} u_7 v_3 + u_5 \right) \\
    \nonumber \frac{v_4}{d\xi} &= \varepsilon_1 \frac{a_0}{z_0} \left( - \bar{\eta} \frac{k_{+4}}{k_{-4}} u_3 v_4 - \bar{\eta} \frac{k_{+4}}{k_{-4}} u_7 v_4 + u_{10} + u_8 \right).
\end{align}
The alternate form giving the functions $f, \; g,$ and $h$ is provided in the supplement.

\subsubsection{Super slow changes in substrate and product concentration}

The limit as $\varepsilon_1 \rightarrow 0$ reflects the fact that the concentrations of substrates far exceeds the concentration of enzyme. In this limit, $\mathbf{v}$ is constant. For the discussion that follows, we will approximate $\mathbf{v}(t)=\mathbf{v}(0)$. This approximation means $[ADP]\approx a_0$, $[PG]\approx 0$, $[ATP]\approx 0$, and $[TZ]\approx z_0$ in the original model. In our dimensionless system this gives $v_1 = 1$, $v_2 = 0$, $v_3 = 0$ and $v_4 = 1$. This elimination of the super slow dynamics simplifies the system to the remaining 8 variables for the enzyme complexes, $\mathbf{u}$, and makes the dissociation of products $v_2$ and $v_3$ in the steps $u_5 \rightarrow u_7$, $u_7 \rightarrow u_1$ and $u_8 \rightarrow u_9$ non-reversible. This change updates the system to be $\frac{d\mathbf{u}}{d\xi} = f(\mathbf{u}, \mathbf{v}(0))+\frac{1}{\varepsilon_2}g(\mathbf{u}, \mathbf{v}(0))$ where 
\begin{equation*}
f(\mathbf{u}, \mathbf{v}(0)) =  \begin{bmatrix}
    k_{-3} u_7 \\
    0 \\
    - k_{+5} u_4 + k_{-5} u_5 \\
    - k_{-5} u_5 + k_{+5} u_4 \\
    - k_{-3} u_7 \\
    - k_{-2} u_8 \\
    k_{-2} u_8 \\
    0
\end{bmatrix} \text{ and } g(\mathbf{u}, \mathbf{v}(0)) = \begin{bmatrix}
    - \frac{k_{+3}}{k_{+4}} u_1 + \frac{k_{+3}}{k_{+4}} u_3 \\
    - \frac{k_{-3}}{k_{+4}} u_3 - u_3 + \frac{k_{+3}}{k_{+4}} u_1 + \frac{k_{-4}}{k_{+4}} u_4 - \bar{\eta} u_3 + \frac{k_{-4}}{k_{+4}} u_{10} \\
    - \frac{k_{-4}}{k_{+4}} u_4 + u_3 \\
    - \frac{k_{-4}}{k_{+4}} u_5 \\
    \frac{k_{-4}}{k_{+4}} u_5 - \bar{\eta} u_7 + \frac{k_{-4}}{k_{+4}} u_8 \\
    - \frac{k_{-4}}{k_{+4}} u_8 + \bar{\eta} u_7 \\
    - \frac{k_{+2}}{k_{+4}} u_9 + \frac{k_{-2}}{k_{+4}} u_{10} \\
    - \frac{k_{-2}}{k_{+4}} u_{10} - \frac{k_{-4}}{k_{+4}} u_{10} + \frac{k_{+2}}{k_{+4}} u_9 + \bar{\eta} u_3
\end{bmatrix}.
\end{equation*}
\subsubsection{Equilibrium approximations}

\begin{figure}
    \centering
    \includegraphics[width=0.5\linewidth]{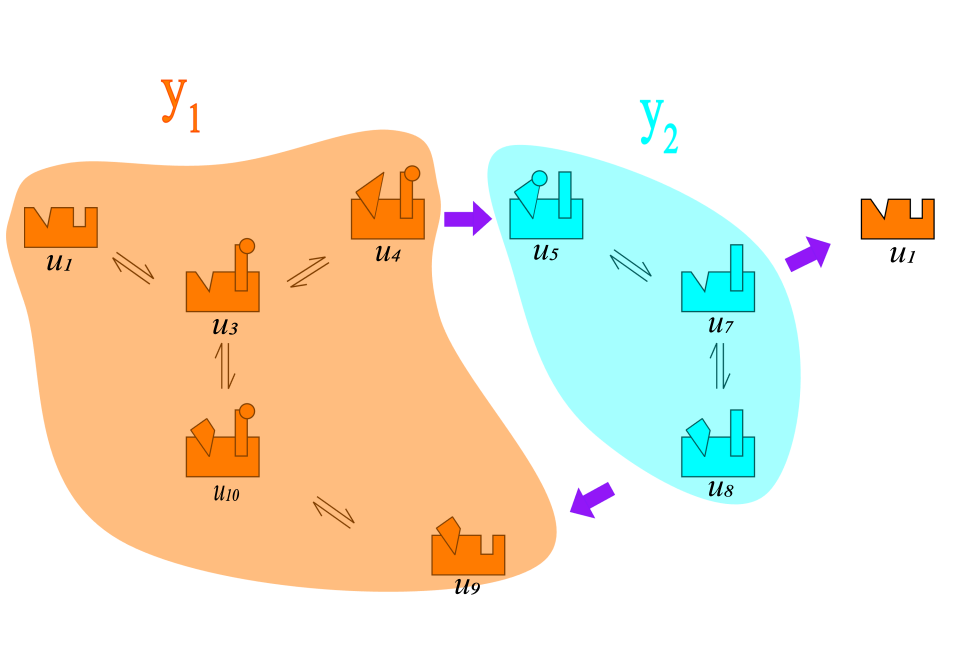}
   \caption{Grouping of the complexes which approach an equilibrium on the fast time scale. In cyan are the complexes which include $v_2$ and make up $y_2$. The complexes that do not include $v_2$ make up $y_1$ and are shown in orange. The three reactions on the slow time scale are shown in purple.}
    \label{Grouping}
    \vspace{-20pt}
\end{figure}

Continuing our reduction to the slow time scale, we take the limit as $\varepsilon_2 \rightarrow 0$. This gives us algebraic constraints $g(\mathbf{u},\mathbf{v}(0))=0$ which define a slow manifold. Restriction to the slow manifold can be viewed as a collection of equilibrium approximations on the fast reactions. 
\begin{align} \label{Equilibrium Approximations}
    \nonumber u_1 &= \frac{k_{-3}}{k_{+3}} u_3 \\
    \nonumber u_3 &= \frac{k_{-4}}{k_{+4}} u_4 \\
    \nonumber u_5 &= 0 \\
    u_8 &= \bar{\eta} \frac{k_{-4}}{k_{+4}} u_7 \\
    \nonumber u_9 &= \frac{k_{-2}}{k_{+2}} u_{10} \\
    \nonumber u_{10} &= \bar{\eta} \frac{k_{+4}}{k_{-4}} u_3
\end{align}
Grouping together the sets of complexes in fast equilibrium we obtain new variables $y_1$ and $y_2$ which give a parametrization of the slow manifold
\begin{eqnarray*}
y_1&=&u_1+u_3+u_4+u_9+u_{10}\\
y_2&=&u_5+u_7+u_8.
\end{eqnarray*}
In Figure \ref{Grouping}, we see that after the grouping into $y_1$ and $y_2$, we have transitions between groups only with our three slow reactions. These three slow reactions, the phosphotransfer, $u_4\rightarrow u_5$, the release of $v_2$ from $u_7$, $u_7\rightarrow u_1$, and the release of $v_2$ from  $u_8$, $u_8 \rightarrow u_9$, are shown in purple. Note that these three reactions are irreversible. $u_7\rightarrow u_1$ and $u_8 \rightarrow u_9$ are irreversible because $v_2=0$ and $u_4\rightarrow u_5$ is irreversible because $u_5=0$. Simplifying, we have
\begin{equation*}
\frac{dy_1}{d \xi}=- \frac{dy_2}{d\xi}=-k_{+5}u_4+k_{-3}u_7 +k_{-2}u_8.
\end{equation*}

The rates of these three remaining reactions can be found by using our algebraic equilibrium approximations from equation \ref{Equilibrium Approximations} to write each complex as a proportion of either $y_1$ or $y_2$. In particular we are interested in the rates of the three slow reactions so we compute $u_4=\gamma y_1$, where 
\begin{equation*}
\gamma=\frac{\sigma \lambda}{\sigma \lambda + \lambda + (1 + \lambda) \tau + \rho \mu}
\end{equation*}
is the proportion of $y_1$ that is $u_4$. In this expression, we have introduced new dimensionless parameters for clarity. The scaled total concentration of ADP and ATP is $\sigma = \frac{k_{+4}}{k_{-4}}$ or in the original parameters $\sigma=\frac{d_+ a_0}{d_-}$. The scaled total concentration of the substrate BPG is $\lambda=\frac{k_{+2}}{k_{-2}}$ or in the original parameters $\lambda = \frac{b_+ p_0}{b_-}$.  We therefore think of $\sigma$ and $\lambda$ as our scaled substrates. Similarly the scaled TZ dose is $\tau=\frac{\bar{\eta}k_{+4}}{k_{-4}}$ or in the original parameters $\tau = \frac{\eta d_+ {z_0}}{d_-} $. The ratio $\mu=\frac{k_{-3}}{k_{-2}}$, or in the original parameters
$\mu = \frac{c_-}{b_-}$, is the ratio of dissociation rates of PG. This is the primary parameter which was explored in the original model \cite{Riley}. Two other ratios of underlying rate parameters will be needed. $\rho = \frac{k_{+2}}{k_{+3}} =\frac{b_+}{c_+} $ captures the extent to which the binding of BPG is reversible. Finally, $\omega=k_{-2}= \frac{b_-}{k_+}$ compares the rates of dissociation of $v_2$ with the central phosphotransfer step. Together, $\omega$ and $\mu$ let us compare the three key reaction rates of the slow timescale. These new parameters are summarized for reference in Table \ref{tab:AllGreek}. 
    
Returning to our equilibrium approximations, now for the reactions in $y_2$, we have $u_7=\frac{1}{1 + \tau} y_2$ and $u_8 =\frac{\tau}{1 + \tau} y_2$. The rate of return from $y_2$ to $y_1$ is $k_{-3} u_7 + k_{-2} u_8=\beta y_2$ where $\beta$ is 
\begin{equation*}
\beta = \frac{\omega(\mu + \tau)}{1 + \tau}.
\end{equation*}
\begin{table}[htbp]
  \centering
        \begin{tabularx}{0.75\textwidth} { 
          | >{\raggedright\arraybackslash}X 
          | >{\centering\arraybackslash}X | >{\centering\arraybackslash}X | }
         \hline
         Parameters & Dimensionless Rate Parameters & Original Rate Parameters \\
         \hline
         $\sigma$ & $\frac{k_{+4}}{k_{-4}}$ & $\frac{d_+ a_0}{d_-}$\\
         \hline
         $\lambda$ & $\frac{k_{+2}}{k_{-2}}$ & $\frac{b_+ p_0}{b_-}$\\
         \hline
         $\tau$ & $\frac{\bar{\eta} k_{+4}}{k_{-4}}$ & $\frac{\eta d_+ z_0}{d_-}$\\
         \hline
         $\mu$ & $\frac{k_{-3}}{k_{-2}}$ & $\frac{c_-}{b_-}$\\
         \hline
         $\rho$ & $\frac{k_{+2}}{k_{+3}}$ & $\frac{b_+}{c_+}$\\
         \hline
         $\omega$ & $k_{-2}$ & $\frac{b_-}{k_+}$\\
         \hline
         \end{tabularx}
         \caption{New dimensionless parameters were chosen to allow easier readability of the reaction rate equation and subsequent results. This table shows the new parameters written in terms of the parameters from the dimensionless model and from the original model.}
         \label{tab:AllGreek}
   \end{table}

The system for $y_1$ and $y_2$ can now be written
\begin{eqnarray*}
    \frac{d y_1}{d\xi} &=& -\gamma y_1 + \beta y_2\\
\frac{d y_2}{d\xi} &=&\gamma y_1 - \beta y_2\\
\end{eqnarray*}
Since $y_1+y_2=1$ is conserved, we have simply $\frac{d y_1}{d \xi}=-\gamma y_1 +\beta(1-y_1).$

We observe that $\gamma$ and $\beta$ are functions of the kinetic parameters and the substrate and drug concentrations.

\subsubsection{Quasi-Steady-State Approximation for the Slow Time Scale}

Our goal is to understand the reaction rate when the slow time scale has reached quasi-steady-state but we have not yet allowed the super-slow variables to evolve. We therefore make one final quasi-steady-state approximation. We estimate $\frac{d y_1}{d \xi}=0$ and we find that the rate of the central phosphotransfer reaction is given by
\begin{equation*}
v=\gamma y_1 = \frac{\gamma \beta}{\gamma +\beta}.
\end{equation*}
This expression approximates the overall reaction rate in the steady state of the slow timescale but before any depletion of substrates or accumulation of products on the super-slow timescale. Thus, it is an approximation of the reaction rate depicted in Figure \ref{fig:ATPProductionModels} and in the simulations in \cite{Riley} of the original model.

Using a combination of quasi-steady-state and equilibrium approximations, we have derived an approximation of the reaction rate as a function of dimensionless kinetic parameters and scaled concentrations of reactants. Our reaction rate equation is given by
\begin{equation}\label{veqn}
    v(\sigma,\lambda,\tau)=\frac{\omega \sigma \lambda (\mu + \tau)}{\sigma \lambda (1 + \tau) + \omega (\sigma \lambda + \lambda + (1 + \lambda) \tau + \rho \mu) (\mu + \tau)}.
\end{equation}
\section{Qualitative Behavior of the Reaction Rate Equation}

\begin{figure}
    \centering
    \includegraphics[width=0.9\linewidth]{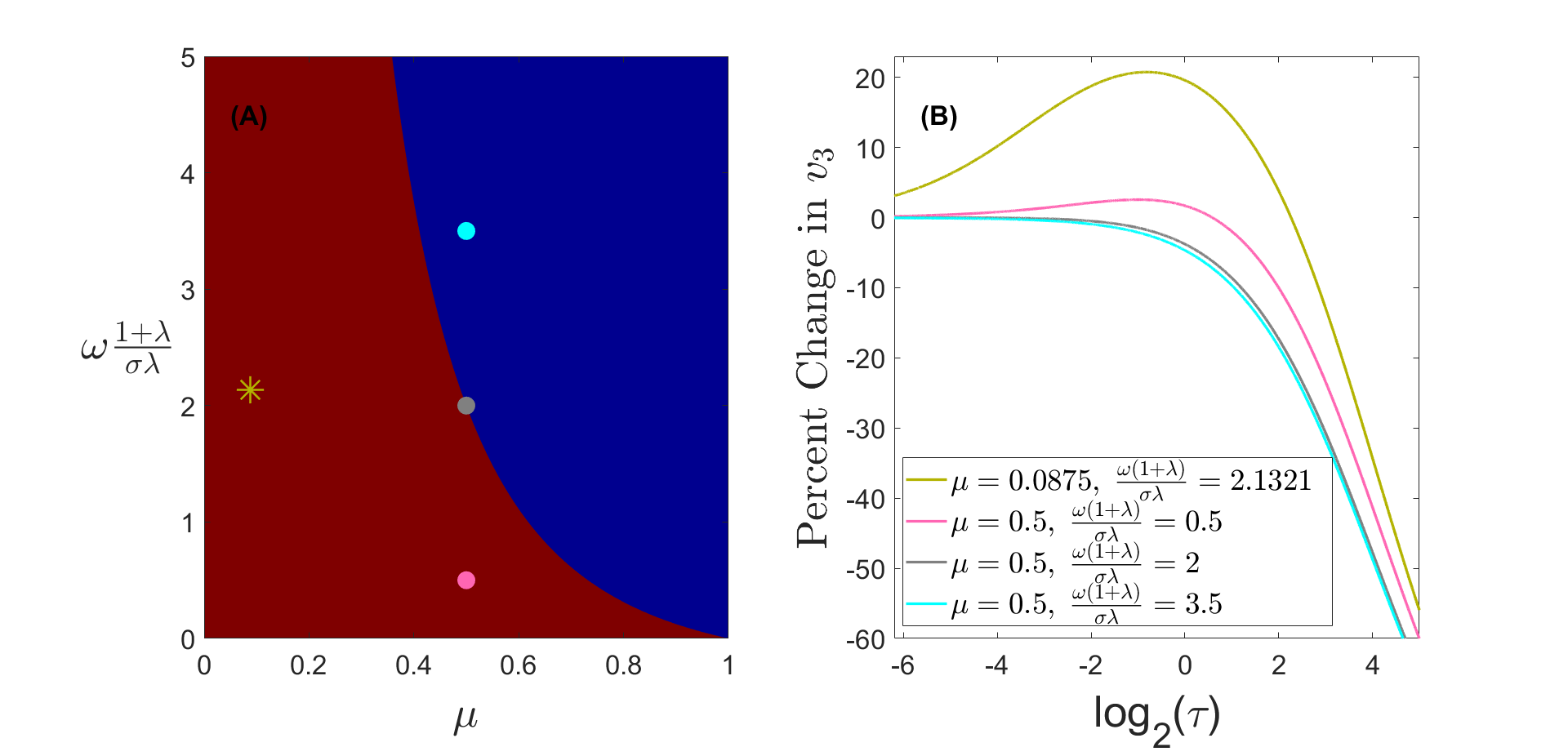}
    \vspace{-5pt}
    
   \caption{Region allowing CIS in the reaction rate equation. Panel A shows the region of parameters where CIS occurs. Points in the red region exhibit CIS whereas points in the blue region exhibit only inhibition. Four representative examples are indicated. The yellow and pink are in the CIS region. The yellow star corresponds to the original parameters. The gray is on the boundary between regions. The cyan is in the region of inhibition only. Panel B shows dose response curves for these four representative parameter sets. In all four cases, the substrate concentrations $\lambda$ and $\sigma$ use the original parameters, and only the value of $\omega$ is varied.}
   \label{fig:ReducedAnalysis}
\end{figure}

The reduction in Section 2 resulted in a single expression for the reaction rate given in equation \ref{veqn}. This section contains analysis of the reaction rate equation and its consequences for our understanding of the CIS mechanism. We derive the parameter regime where CIS may occur, the $v_4$ dose which results in maximal reaction rate, and the maximal $v_4$ dose which generates stimulation. Finally we compare this reaction rate equation to classic reaction rate equations for enzyme kinetics.

\subsection{Region of parameters space that allows stimulation}
 The reaction rate in equation \ref{veqn} is given as a rational function. It has a Monod form when viewed as a function of either substrate, $\sigma$ or $\lambda$. As a function of drug dose, $\tau$, the numerator grows linearly while the denominator grows quadratically. This is consistent with the biphasic dose response in the full model. To compute the range of parameters that allow CIS, we note that because all parameters are non-negative, $v(\sigma,\lambda,\tau)\ge0$, and $\lim_{\tau \rightarrow \infty}v(\sigma,\lambda,\tau)=0$. To show a biphasic dose response, we need only show that $\frac{dv}{d\tau} \left. \right|_{\tau = 0}>0$. A straightforward calculation gives us the condition 
\begin{equation}\label{MuInequality}
\mu^2 + \frac{\sigma \lambda}{\omega (1 + \lambda)} \mu - \frac{\sigma \lambda}{\omega (1 + \lambda)} < 0.
\end{equation}
This relationship is shown in Figure \ref{fig:ReducedAnalysis}. The red area is combinations of parameters that exhibit CIS, while those in blue exhibit only inhibition. This elucidates the observation in \cite{Riley} of the importance of $\mu$, the ratio of the rates of dissociation of $v_2$. Specifically if $\mu$ is small, we expect to have CIS for a broad range of values of $\omega$ and a broad range of substrate concentrations $\sigma$ and $\lambda$. The original parameters (those used in Figures \ref{fig:threescales}, \ref{fig:FluxComparison} and \ref{fig:ATPProductionModels}, and adapted from \cite{Riley}) are shown with a yellow dot. The other three dots have a larger value of $\mu$ and are chosen as representative examples of CIS (pink), inhibition only (cyan) and the boundary between regions (gray). The dose response for these four parameter sets are shows in Panel B to illustrate the transition from CIS to inhibition only. Equation \ref{MuInequality} also highlights the importance of the combination of parameters $\omega \frac{1+\lambda}{\sigma\lambda}$. In panel B we held the substrate levels, $\sigma$ and $\lambda$, at the original values and adjusted $\omega$ to obtain the desired combinations. Recall that in the original system $\omega=b_{-}/k_+$ is the ratio of the rate of $v_2$ release from $u_8$ to the rate of the phosphotransfer $u_4 \rightarrow u_5$. Together $\mu$ and $\omega$ capture the relationship among the three reactions on the slow time scale. 

\subsection{Maximal rate and maximal dose}
Next, we note that for $\tau=0$, the reaction rate, $v$, is increasing as a function of both $\sigma$ and $\lambda$. That is, with no drug, $v_4$, the reaction rate is larger with more substrate and has a maximum rate of 
\begin{equation}\label{vmaxnoTZ}
\bar{v}=\frac{\omega \mu}{\omega \mu+1}.
\end{equation}
Returning to the original parameters, this maximum rate of the TZ-free reaction is $\frac{c_- k_+}{c_- \; + \; k_+} e_0$ emphasizing the important interplay between the two slow steps, the phosphotransfer step with rate $k_+$ and the dissociation of PG from E$\cdot$PG with rate $c_-$. 

Two special doses are clinically important. The optimal dose which results in the highest reaction rate and the maximum stimulating dose which is the threshold between stimulation and inhibition. If a drug, such as TZ, is being administered to increase the enzyme's activity, one would like to stay near the optimal dose while ensuring that you do not exceed the maximum stimulating dose. Similarly, if one is attempting to create a novel drug based on this CIS mechanism, higher reaction rate at the optimal dose and higher maximum stimulating dose would be preferred. 

The optimal dose is the value of $\tau$ which gives the largest reaction rate. This optimal dose can be computed from equation \ref{veqn} by setting $\frac{dv}{d\tau}=0$ and solving for $\tau$. The optimal dose is the positive solution,  
\begin{equation}\label{OptimalDose}
\tau=- \mu + \sqrt{\frac{\sigma \lambda (1 - \mu)}{\omega (1 + \lambda)}}.\end{equation}
Note that this value of $\tau$ will be real and positive whenever equation \ref{MuInequality} is satisfied.  This optimal value for $\tau$ is shown in Figure \ref{fig:ParameterRegimeOriginal} B as a white curve. The proportional increase in reaction rate, $v(\sigma,\lambda,\tau)/v(\sigma,\lambda,0)$, at this optimal dose can be written as a function of the parameters $\sigma, \lambda, \mu, \omega,$ and $\rho$. The resulting expression is unwieldy but useful. For example, for reasonable choices of the other parameters one can show that the proportional increase in reaction rate at the optimal dose increases as $\mu$ decreases. 

The maximum stimulating dose can be computed by finding the value of $\tau$ which produces the same value for $v$ as $\tau=0$. That is, setting $v(\sigma,\lambda,\tau)=v(\sigma,\lambda,0)$, we have 
\begin{equation}\label{maxStimDose}
\tau=\frac{1-\mu}{\mu} \frac{\sigma\lambda}{\omega(1+\lambda)}-\mu.
\end{equation}
Again, this will give a positive value for $\tau$ whenever equation \ref{MuInequality} is satisfied. This value for $\tau$ is shown in Figure \ref{fig:ParameterRegimeOriginal} B as a black curve. This maximum dose is a decreasing function of both $\mu$ and $\omega$ suggesting that drug development will work best when both $\mu$ and $\omega$ are small. Note that the optimal dose and the maximum stimulating dose given are scaled. So, in addition to the parameters listed here, the optimum and maximum doses in practice will be proportional to the dissociation constant for the drug.  

\subsection{Comparison to Classic Reaction Rate Models}

\begin{figure}
    \centering
    \includegraphics[width=0.9\linewidth]{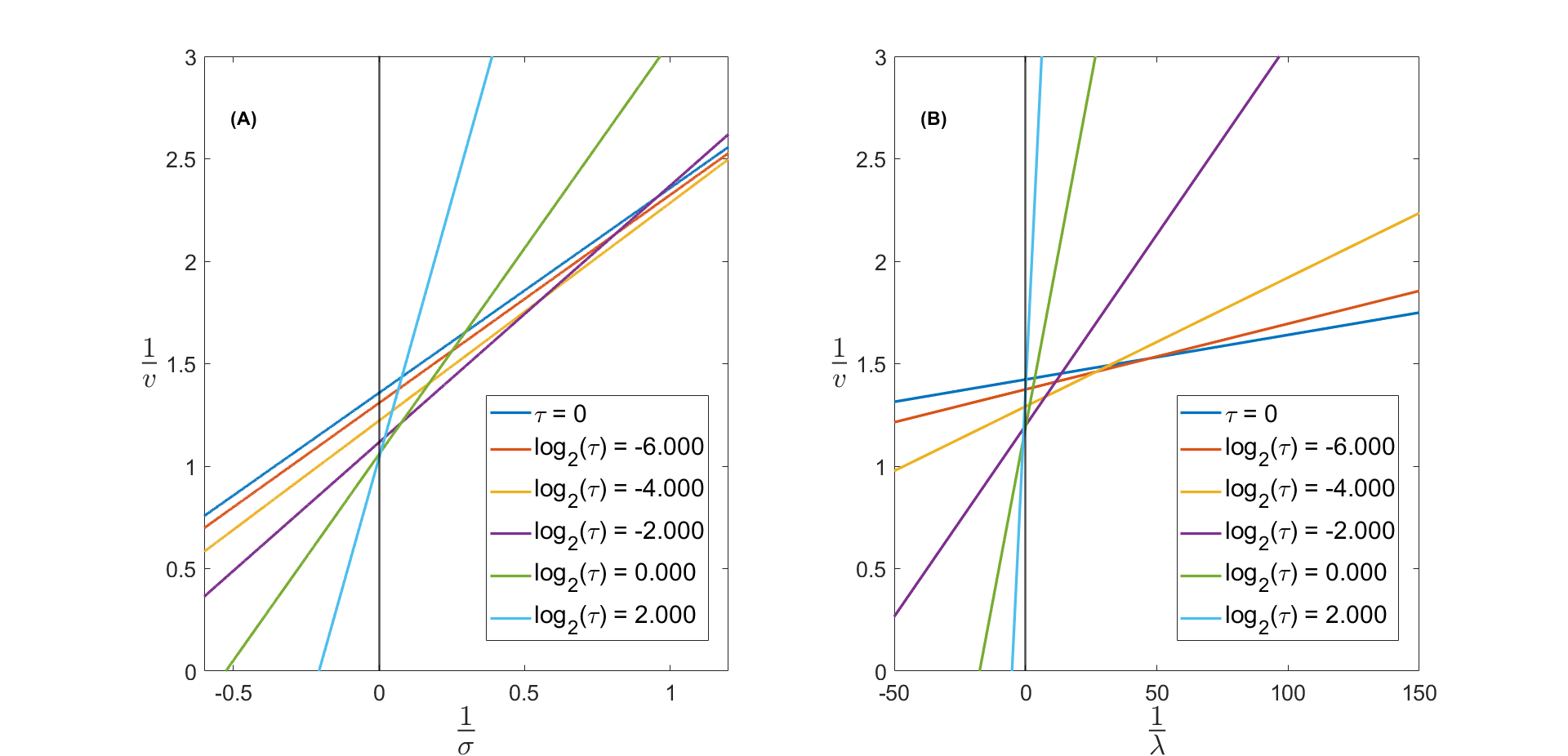}
    \vspace{-5pt}
    
   \caption{Lineweaver Burk plots. As the dose of TZ, $\tau$, is increased, the lines shift lower vertically and then rotate around their $1/v$ intercept. This downshift represents an increase in $v_{max}$. The increase in slope corresponds to an increase in the Michaelis constant. At low and moderate doses, the drug increases $v_{max}$ while at higher doses, the drug is behaving as a classic competitive inhibitor. }
   \label{fig:LWB}
\end{figure}

The rate equation is a rational function which is linear in dose, $\tau$, in the numerator and quadratic in the denominator. This means that for the parameters described in equation \ref{MuInequality}, the rate first increases with $\tau$ but for large values of $\tau$, the rate will approach zero. This rational form is similar to the classic Haldane model for substrate inhibition which is proportional to the substrate in the numerator but has a quadratic in the denominator. Thus, that model also exhibits biphasic dose response, but differs in that the response is to the substrate only and that the reaction has zero rate in the absence of that substrate. 

We turn next to a comparison with the standard two substrate Michaelis-Menton rate equation. First, this formula can be written in a Michaelis-Menton (Monod function) form as a function of either substrate, $\lambda$ or $\sigma$.
\begin{align*}
v &= \frac{\omega \lambda (\mu + \tau) \cdot \sigma}{(\lambda (1 + \tau) + \omega \lambda (\mu + \tau)) \cdot \sigma \; \;  +  \;  \; \omega (\lambda + (1 + \lambda) \tau + \rho \mu)(\mu + \tau)} \\
v &= \frac{\omega \sigma (\mu + \tau) \cdot \lambda}{(\sigma (1 + \tau) + \omega \sigma (\mu + \tau) + \omega \mu (1 + \tau) + \omega \tau (1 + \tau)) \cdot \lambda \; \; + \; \; \omega (\tau + \rho \mu) (\mu + \tau)}
\end{align*}
The Michaelis-Menton form can be seen in the linear Lineweaver-Burke plots in Figure \ref{fig:LWB}. Panel A shows $\frac{1}{v}$ as a function of $\frac{1}{\sigma}$ and panel B shows $\frac{1}{v}$ as a function of $\frac{1}{\lambda}$. However, they are not combined in a standard two substrate Michaelis-Menton expression, $\frac{v_{max}s_1 s_2}{(s_1+K_1)(s_2+K_2)}$, because there is no term in the denominator which is proportional to $\sigma$ only. This form is, however, consistent with the equilibrium approximation for an ordered mechanism, $\frac{v_{max}s_1 s_2}{s_1 s_2 + K_2 s_1 +K_1 K_2}$.
We next rewrite the reaction rate given in equation \ref{veqn} in a form reminiscent of this standard ordered mechanism rate equation where the quantities $v_{max}^{\text{eff}}$, $K_{\sigma}^{\text{eff}}$, and $K_{\lambda}^{\text{eff}}$ are written as functions of $\tau$.
\begin{equation*}
v=\frac{v_{max}^{\text{eff}}(\tau) \sigma \lambda }{\sigma\lambda +K_{\sigma}^{\text{eff}}(\tau)\lambda+K_{\sigma}^{\text{eff}}(\tau)K_{\lambda}^{\text{eff}}(\tau)}
\end{equation*}
where
\begin{equation*}
v_{max}^{\text{eff}}(\tau)=\frac{\omega (\mu +\tau)}{\omega(\mu + \tau)+(1+\tau)}, \; \;  K_{\sigma}^{\text{eff}}(\tau)=\frac{\omega (\mu +\tau)(1+\tau)}{\omega(\mu + \tau)+(1+\tau)}
\text{ and }
K_{\lambda}^{\text{eff}}(\tau)=\frac{(\rho \mu+\tau)}{(1+\tau)}.
\end{equation*}
Here, we use the $\text{eff}$ superscript to indicate that these are the ``effective" quantities taking into account the impact of the drug, $v_4$, and the interplay of the relevant dimensionless parameters. An exploration of the leading order dose response for small, medium, and large doses illustrates the main features. 
For very small doses, $\tau <<\mu$, we have $v_{max}^{\text{eff}}\approx K_{\sigma}^{\text{eff}}\approx \bar{v}$ and $ K_{\lambda}^{\text{eff}}\approx \rho \mu$ where $\bar{v}$ is the no-drug maximum reaction rate defined in equation \ref{vmaxnoTZ}. Together, this gives a reaction rate of  
\begin{equation*}
    v\approx\frac{\bar{v} \sigma \lambda}{\sigma \lambda+\bar{v}\lambda+\bar{v}\rho \mu }.
\end{equation*}
For very large doses, $\tau >> 1$, we have $v_{max}^{\text{eff}}(\tau)$ is approximated by its maximum value 
\begin{equation}\label{vhat}
\hat{v}=\frac{\omega}{\omega+1}.\end{equation}
  $K_{\sigma}^{\text{eff}}$ is approximately $\hat{v} \tau$ and $K_{\lambda}^{\text{eff}}$ is approximately 1. This gives a form which mirrors the classic competitive inhibition model with linear growth in one of the $K$ values, 
  \begin{equation*}
  v\approx\frac{\hat{v}\sigma \lambda}{\sigma\lambda+\hat{v}\tau\lambda +\hat{v}\tau}.
  \end{equation*}
This classic competitive inhibition behavior is observed in higher doses in the Lineweaver-Burke plots (Figure \ref{fig:LWB}) as an increasing slope, but unchanged $1/v$ intercept. 

For moderate values of $\tau$, $\mu<<\tau<<1$, we have $v_{max}^{\text{eff}}$ and $K_{\sigma}^{\text{eff}}$ both approximately $\omega \tau$ and $K_{\lambda}^{\text{eff}} \approx \tau$ capturing the crucial quadratic term in the denominator. In Figure \ref{fig:LWB}A we see that for these moderate doses, while both $v_{max}^{\text{eff}}$ and $K_{\sigma}^{\text{eff}}$ are increasing, their ratio is relatively constant. This ratio is the slope of the Lineweaver-Burke plot in Panel A, therefore, the lines shift downward before they begin to rotate. In comparison, the slope $K_{\lambda}^{\text{eff}} /v_{max}^{\text{eff}}$ in panel B is growing throughout the range of doses. The approximate reaction rate equation in this middle range of doses is
\begin{equation*}
v\approx \frac{\omega \tau \sigma \lambda}{\sigma\lambda + \omega \tau \lambda + \omega \tau^2}.
\end{equation*}
Finally, we see that the functions $v_{max}^{\text{eff}}$, and $K_{\sigma}^{\text{eff}}$ are similar so we rewrite the reaction rate in Equation \ref{veqn} in yet another form, 
\begin{equation*}
v=\frac{m(\tau) \sigma \lambda }{\sigma\lambda +m(\tau)(1+\tau)\lambda+m(\tau)(\rho \mu+\tau)}
\end{equation*}
where the function $m$ is given by
\begin{equation*}
m(\tau)=\frac{\omega (\mu +\tau)}{\omega(\mu + \tau)+(1+\tau)}.
\end{equation*}
Note that $m(\tau)$ can also be written $\frac{\beta}{1+\beta}$ indicating the importance of $\beta$, the total rate at which the $v_2$ is released. The Michaelis constant for $\sigma$ can be written in terms of $m$ as $K_{\sigma}^{\text{eff}}=m(\tau)(1+\tau)$. Fundamentally, we wish to understand how this function $m(\tau)$ depends on the dose $\tau$. When there is no drug, we have $m(0)=\bar{v}$ where $\bar{v}$ is the no-dose maximum reaction rate defined in equaiton \ref{vmaxnoTZ}. Rewriting $m(\tau)$ in terms of the proportional increase relative to $\bar{v}$ gives a function of the form
\begin{equation*}
m(\tau)=\bar{v}\left(1+A \frac{\tau}{K_{\tau}+\tau}\right).
\end{equation*}
The increase in $m(\tau)$ relative to the no-dose value is also a Monod function which increases and asymptotes to a maximum value of $\hat{v}$ defined in equation \ref{vhat}. The proportional increase in $m(\tau)$ is given by
\begin{equation*}
A=\frac{\hat{v}-\bar{v}}{\bar{v}}=\frac{1-\mu}{\mu(\omega+1)}.
\end{equation*}
The quantity $A$ gives the maximal proportional increase in $m$ and therefore also in $v_{max}^{\text{eff}}$. $A$ is positive when $\mu<1$ and is largest when $\mu$ is small. This confirms our intuition that the reaction rate can be increased when $\mu<1$, or in the original PGK1 system, when the dissociation of PG is faster when bound to TZ. The constant $K_{\tau}= \frac{\hat{v}\mu}{\bar{v}}=\frac{\omega \mu+1}{\omega+1}$ can be viewed as giving the essential scaling of the dose $\tau$. Smaller values of $\mu$ lead to smaller values of $K_{\tau}$ which in turn leads to increase in $m(\tau)$ for smaller values of $\tau$

These changes in effective maximum rate and Michaelis constants can be seen in the Lineweaver-Burke plots in Figure \ref{fig:LWB}. In both panels, we see the lines shift down and then rotate. At low doses of $\tau$ the downward shift indicates an increase in $v_{max}^{\text{eff}}$. As the dose continues to increase, the $1/v$ intercept changes very little as the slope increases, consistent with competitive inhibition. 

In summary, the effect of TZ is to increase $v_{max}^{\text{eff}}$ at low doses. However, the effect on $v_{max}^{\text{eff}}$ levels out at $\hat{v}$ for higher doses. Similarly, $K_{\lambda}^{\text{eff}}$ approaches one for large doses. At the same time, the effective Michaelis constant, $K_{\sigma}^{\text{eff}}$ is a linear function times $m(\tau)$, so even as $m(\tau)$ approaches its maximum, this Michaelis constant continues to grow linearly with dose. This is consistent with the classic competitive inhibitor model. This reduction of the original PGK1 model captures the qualitative biphasic dose response with respect to TZ, the standard Michaelis-Menton dose response to both substrates, and sheds light on the fundamental mechanism of CIS. In this model, we see that the response to high doses of TZ is dominated by the linear increase in the $K_{\sigma}^{\text{eff}}$. On the other hand, the stimulation at low doses is facilitated by the increase in $v_{max}^{\text{eff}}$.

\section{Comparison to Full Model}

In this section, we compare simulations of the dimensionless model (equation \ref{dimensionless}) with the reaction rate equation (equation \ref{veqn}). In section 4.1 we compare the region of parameter space which exhibits CIS and in 4.2 we return to the relevance of the parameter $\mu$. In Figures \ref{fig:OmegaVsMuHeat} and \ref{fig:ParameterRegimeOriginal} we compare heatmaps of the reaction rate relative to the zero dose reaction rate. 

\subsection{Biphasic Dose Response}

\begin{figure}
    \centering
    \includegraphics[width=1.0\linewidth]{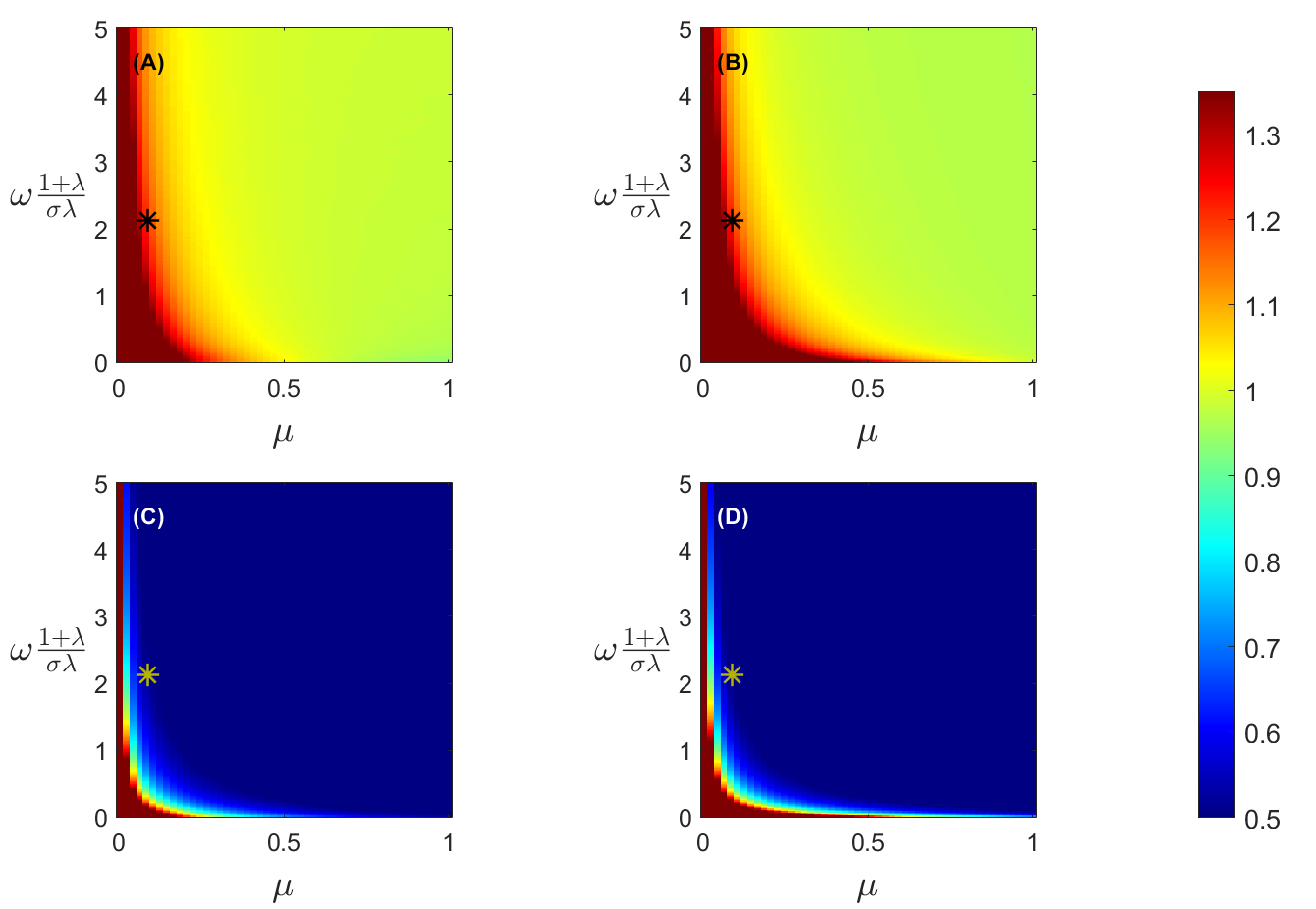}
    \vspace{-5pt}
    
   \caption{The biphasic response in the dimensionless model and the reaction rate equation. As in Figure \ref{fig:ReducedAnalysis}, only $\mu$ and $\omega$ are being changed within each panel and the substrate concentrations, $\sigma$ and $\lambda$, are held constant at the original parameters. Panels A and B use a moderate dose,$z_0 = 50 n$M ($\tau$ is scaled $z_0$). This drug dose has many parameter combinations that stimulate production and all other combinations have little effect. In contrast, panels C and D use a high dose, $z_0 = 2.5 \mu$M. At this dose, most parameter values lead to inhibition. Panels A and C show the ratio of $v_3$ generated by the dimensionless model to the no dose value of $v_3$. All values were measured at $\xi = 300$. Panels B and D heatmaps were generated by calculating the ratio of the reaction rate compared to the no-dose reaction rate when $\tau = 0$. In all four panels, the star represents the original parameter values.  Together, these panels demonstrate that the reaction rate model captures the behavior of the dimensionless model.}
   \label{fig:OmegaVsMuHeat}
\end{figure}

The biphasic region was observed in Figure \ref{fig:ReducedAnalysis} panel A, but different parameter regimes will produce more intense stimulation or have a stronger inhibitory effect on the production of product. To see how well the reaction rate equation approximates the results of the dimensionless model, we compare heatmaps in Figure \ref{fig:OmegaVsMuHeat}. In all panels, the star shows the original parameter values.

Panels A and B show both models when given the same stimulating dose of TZ. In the dimensionless model, we show the results for $z_0 = 50$nM. This dose is rescaled to $\tau = \frac{\eta d_+}{d_-} z_0$ for the reaction rate equation. The reaction rate in Panel A is computed by running the dimensionless model up to $\xi = 300$ and then plotting the ratio of $v_3$ produced at the this dose divided by $v_3$ produced when there is no TZ present. The relative reaction rate in panel B is computed directly from equation \ref{veqn}, $v(\sigma,\lambda,\tau)/v(\sigma,\lambda,0)$. Red values indicate an increase in reaction rate while blue values indicate a decrease. This is done at differing values of $\mu$ and $\omega$ while keeping $\sigma$ and $\lambda$ constant at the original values. These two panels reveal that at low values of $\mu$, this dose of TZ will be stimulatory. The slight differences are that the reaction rate equation exhibits stimulation for higher values of $\mu$ when $\omega$ is near $0$ and the reaction rate equation has a larger region of light green meaning slight inhibition is occurring.

Panels C and D are generated similarly, but with a high dose, $z_0 = 2.5 \mu$M. These heatmaps are also strikingly similar. In both the dimensionless model and the reaction rate equation, the majority of parameter values slow the reaction significantly. We observe that for this dose of TZ, the only way to increase product production is to have $\mu$ near $0$ or to have small $\mu$ and small $\omega$. The slight difference between the dimensionless model and the reaction rate equation is that again the reduced model shows stimulation for more values of $\mu$ when $\omega$ is near $0$.

Comparing panels A and B at a moderate dose to panels B and C at the high dose illustrates the intrinsic biphasic dose response. In Panels A and B, at a moderate dose, all combinations of parameters produced either stimulation (red) or only mild inhibition (light green). In contrast, Panels C and D at the high dose, exhibit stimulation for small $\mu$ and $\omega$ but inhibition (dark blue) for much of the parameter space. Comparing panels A and C for the dimensionless model to B and D for the reaction rate equation show that the reaction rate equation reproduces the qualitative results of the full model well.

\subsection{\texorpdfstring{Dependence on $\mu$}{Dependence on mu}}

\begin{figure}
    \centering
    \includegraphics[width=0.4\linewidth]{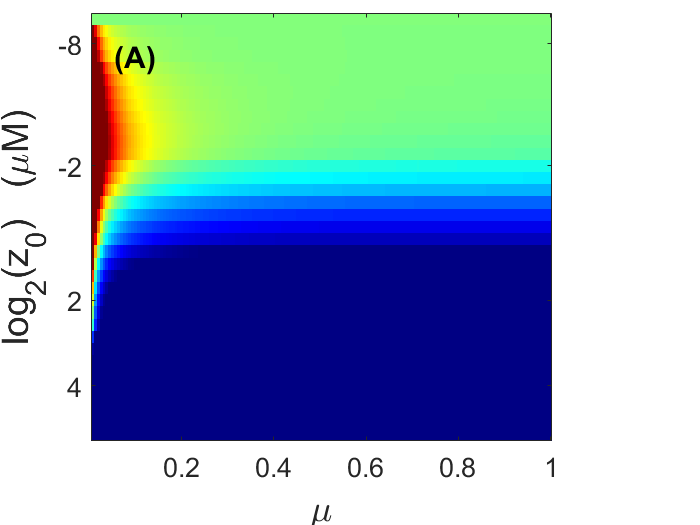}
    \includegraphics[width=0.4\linewidth]{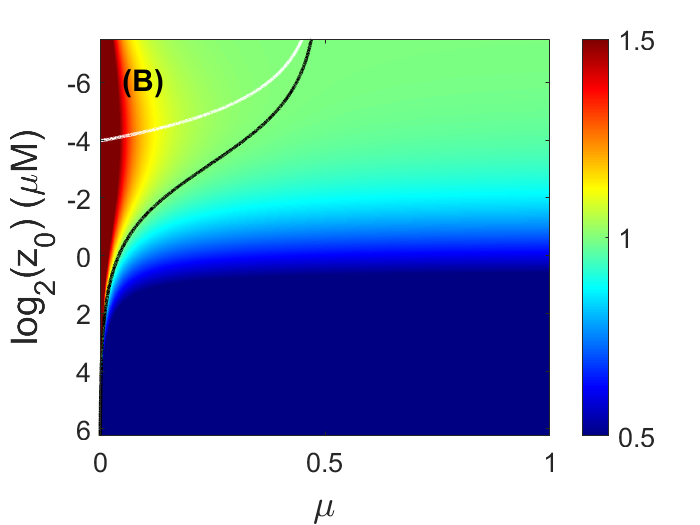}
    \vspace{-5pt}
    
   \caption{Reaction rate as a function of TZ dose. Panel A is computed with simulations of the dimensionless model. For each value of $\mu$, the reaction rate at $\xi=300$ is divided by the reaction rate with the same $\mu$ b ut with no TZ. Red indicates region where TZ increases the reaction rate and blue indicates regions where TZ decreases the reaction rate. This panel replicates the observations in \cite{Riley}. Panel B shows the same proportional change in reaction rate for the reaction rate equation  \ref{veqn}. The curve in white shows the optimal dose which results in the maximum reaction rate. The curve in black shows the maximum stimulating dose which defines the boundary between stimulation and inhibition.}
   \label{fig:ParameterRegimeOriginal}
\end{figure}

In \cite{Riley} we observed that $\mu$ was an important parameter in determining the behavior of the model. The dimensionless model is a rescaling of the original model and inherits this property. Here we examine the effect of $\mu$ in the reaction rate equation. In Figure \ref{fig:ParameterRegimeOriginal}, we once again show heatmaps for the ratio of the reaction rate with TZ versus zero TZ. Green represents when the drug has no effect. Red indicates increased $v_3$ production while blue indicates a decrease. The full dimensionless model in Panel A shows small doses of drug paired with small values of $\mu$ will generate significant increases in $v_3$. However, larger values of $\mu$ decrease the drug's effectiveness no matter the dose. Panel B was generated using the reaction rate equation \ref{veqn} and shows qualitatively similar behavior. One benefit of the reaction rate equation is that the optimal dose (equation \ref{OptimalDose}) and maximal stimulating dose (equation \ref{maxStimDose}) can be explicitly computed. They are represented in the heatmap by the white and black lines, respectively. 

In this section, we have shown that the reaction rate equation captures the essential features of the original model. Shifting our focus, to think of the original model as a general model for competitive inhibition in a two substrate enzyme, we have illustrated the combinations of kinetic parameters which allow for CIS.

\section{Discussion}

We use dynamical systems models, simulation, and fast-slow analysis to tackle the question: {\it  how can a competitive inhibitor {\textbf{stimulate}} enzyme activity in a dose-dependent manner?} We began with a detailed kinetic model of competitive inhibition of PGK1. Using a three timescale reduction, we arrive at a single equation for the reaction rate at the start on the super-slow timescale. This reaction rate equation sheds light on the enzyme assay experiments. It exhibits the biphasic dose response and allows direct computation of optimal doses as a function of underlying rate parameters. By computing kinetic parameters which allow CIS, this model illuminates the mechanism by which TZ stimulates PGK1 activity. 

One possible limitation of the original model is that the reaction rates are chosen to be symmetric (for example, they have the same rate for the dissociation of ATP as those measured for ADP). This simplifies some of the computations. If we disambiguate the parameters in such a way that the same three reactions define the slow timescale, the form of the reduction will be unchanged, but the composite parameters, $\mu$, $\omega$, and $\rho$ will have a less simple form. Further, we conjecture that if the rate of ADP binding or ATP dissociation are on a the slow timescale, the functional form of the reduction will remain. More broadly, we conjecture that if a drug creates a pathway in the enzyme's cycle of reactions that allows the bypass of the rate limiting step, that drug may produce CIS.

The original model gave new insight into TZ's activation of PGK1, but can also be viewed as a general model for an enzyme with two substrates and two products. That is, it can be considered a general model for competitive inhibitors for large classes of enzymes including oxidoreductases, transferases and hydrolases. It predicts that other enzymes with the equivalent of small $\mu$ (where the release of a product is faster in the presence of another substrate or coenzyme) may be targets for activation by a competitive inhibitor. This identification could prove valuable for future drug development since activation is generally more difficult than inhibition. Intuitively, this is because it is more difficult to enhance the function of an enzyme than it is to break or block that function.
One potential challenge for this identification is that the majority of studies report only composite parameters such as $K_{cat}$, $K_m$, and $K_i$. Our estimates of the dimensionless parameters that allow CIS can facilitate the identification of a pool of candidate enzymes. From this pool, we can look for those enzymes that have analogous kinetic parameters in the subset of the parameter space which allows CIS.

Returning to the motivating example of PGK1 and TZ. ATP levels are tightly controlled by regulatory feedback. The functional form for the reaction rate derived here adds to our modeling repertoire and opens the potential to link CIS dynamics for PGK1 into larger metabolic network models. The goal of such models would be to understand how shifting the activity of this enzyme can lead to the substantial changes in ATP levels observed in vitro and in vivo. Such increases could have a therapeutic effect for people with Parkinsons Disease or other diseases with metabolic dysregulation.

\section{Authorship and Contribution Statement}

All authors have made substantial intellectual contributions to the study conception, execution, and design of the work. All authors have read and approved the final manuscript.  In addition, the following contributions occurred:  Conceptualization: Mitchell Riley and Colleen Mitchell; Methodology: Garrett Young and Colleen Mitchell; Formal analysis and investigation: Garrett Young and Colleen Mitchell; Writing - original draft preparation: Colleen Mitchell and Garrett Young; Writing - review and editing: Colleen Mitchell, Garrett Young, Mitchell Riley.

\section{Conflicts of Interest}

The authors have no conflicts of interest.

\section{Data and Code Availability}

The code that supports the findings of this study are openly available in Stimulation-of-Enzymatic-Activity-by-a-Competitive-Inhibitor at DOI: 10.5281/zenodo.16944092.

\section{Funding Statement}

This work was supported by the Iowa Cystic Fibrosis Foundation Research \& Development Program - In Vitro Models Core. Project Number: Stoltz24RO. PD/PI: Stoltz, David SCOR/PPG Director; Welsh, Michael PI.

\bibliographystyle{plain}
\bibliography{main}

\end{document}

\title{Supplement to ``Fast-Slow Analysis of a Model For the Stimulation of Enzymatic Activity by a Competitive Inhibitor''}
\date{}
\maketitle
\thispagestyle{firstpage}

\section{Alternate Notation for the Model}

After removing $u_2$, $u_6$, and $u_1 \leftrightarrow u_9$, the system can be written as:

\[
f(\mathbf{u},\mathbf{v}) =
\begin{bmatrix}
    - k_{+3} u_1 v_2 + k_{-3} u_7 \\
    0 \\
    - k_{+5} u_4 + k_{-5} u_5 \\
    - k_{-5} u_5 + k_{+5} u_4 \\
    - k_{-3} u_7 + k_{+3} u_1 v_2 \\
    - k_{-2} u_8 + k_{+2} u_9 v_2 \\
    - k_{+2} u_9 v_2 + k_{-2} u_8 \\
    0
\end{bmatrix}
\]

\[
g(\mathbf{u}, \mathbf{v}) =
\begin{bmatrix}
    - \frac{k_{+3}}{k_{+4}} u_1 + \frac{k_{-3}}{k_{+4}} u_3 \\
    - \frac{k_{-3}}{k_{+4}} u_3 - u_3 v_1 + \frac{k_{+3}}{k_{+4}} u_1 + \frac{k_{-4}}{k_{+4}} u_4 - \bar{\eta} u_3 v_4 + \frac{k_{-4}}{k_{+4}} u_{10} \\
    - \frac{k_{-4}}{k_{+4}} u_4 + u_3 v_1 \\
    - \frac{k_{-4}}{k_{+4}} u_5 + u_7 v_3 \\
    - u_7 v_3 + \frac{k_{-4}}{k_{+4}} u_5 - \bar{\eta} u_7 v_4 + \frac{k_{-4}}{k_{+4}} u_8 \\
    - \frac{k_{-4}}{k_{+4}} u_8 + \bar{\eta} u_7 v_4 \\
    - \frac{k_{+2}}{k_{+4}} u_9 + \frac{k_{-2}}{k_{+4}} u_{10} \\
    - \frac{k_{-2}}{k_{+4}} u_{10} - \frac{k_{-4}}{k_{+4}} u_{10} + \frac{k_{+2}}{k_{+4}} u_9 + \bar{\eta} u_3 v_4
\end{bmatrix}
\]

\[
h(\mathbf{u},\mathbf{v}) =
\begin{bmatrix}
    - \frac{k_{+4}}{k_{-4}} u_3 v_1 + u_4 \\
    - \frac{a_0}{p_0} \frac{k_{+3}}{k_{-4}} u_1 v_2 + \frac{a_0}{p_0} \frac{k_{-3}}{k_{-4}} u_7 - \frac{a_0}{p_0} \frac{k_{+2}}{k_{-4}} u_9 v_2 + \frac{a_0}{p_0} \frac{k_{-2}}{k_{-4}} u_8 \\
    - \frac{k_{+4}}{k_{-4}} u_7 v_3 + u_5 \\
    - \eta \frac{k_{+4}}{k_{-4}} u_3 v_4 - \eta \frac{k_{+4}}{k_{-4}} u_7 v_4 + \frac{a_0}{z_0} u_{10} + \frac{a_0}{z_0} u_8
\end{bmatrix}
\]

\section{Figures}

\begin{figure}[H]
    \includegraphics[width=1\linewidth]{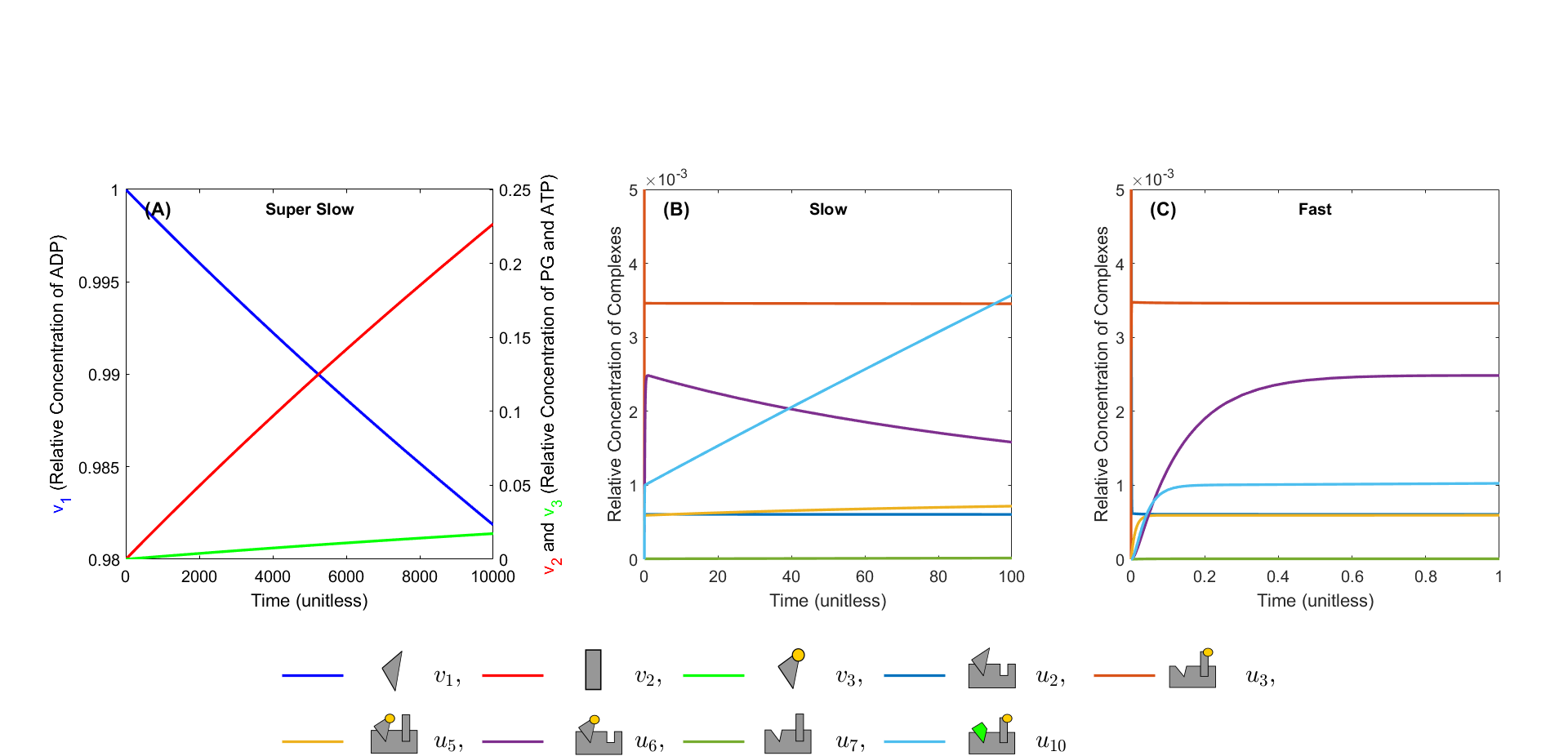}
    \vspace{-15pt}
   \caption{Simulations of the non-dimensionalized model reveal three distinct time scales. This figure was made with a high drug dose ($z_0 = 25\mu$M). Panel (A) shows super-slow changes to substrate and product concentrations over 10,000 units of time. Panel (B) shows changes in the enzyme complexes over 100 time steps. Panel (C) shows the fastest changes over the first unit of time. Each unit of time is $\frac{1}{k_+}=0.2$ seconds.}
   \label{fig:threescaleshighdose}
\end{figure}

\begin{figure}[H]
    \includegraphics[width=1\linewidth]{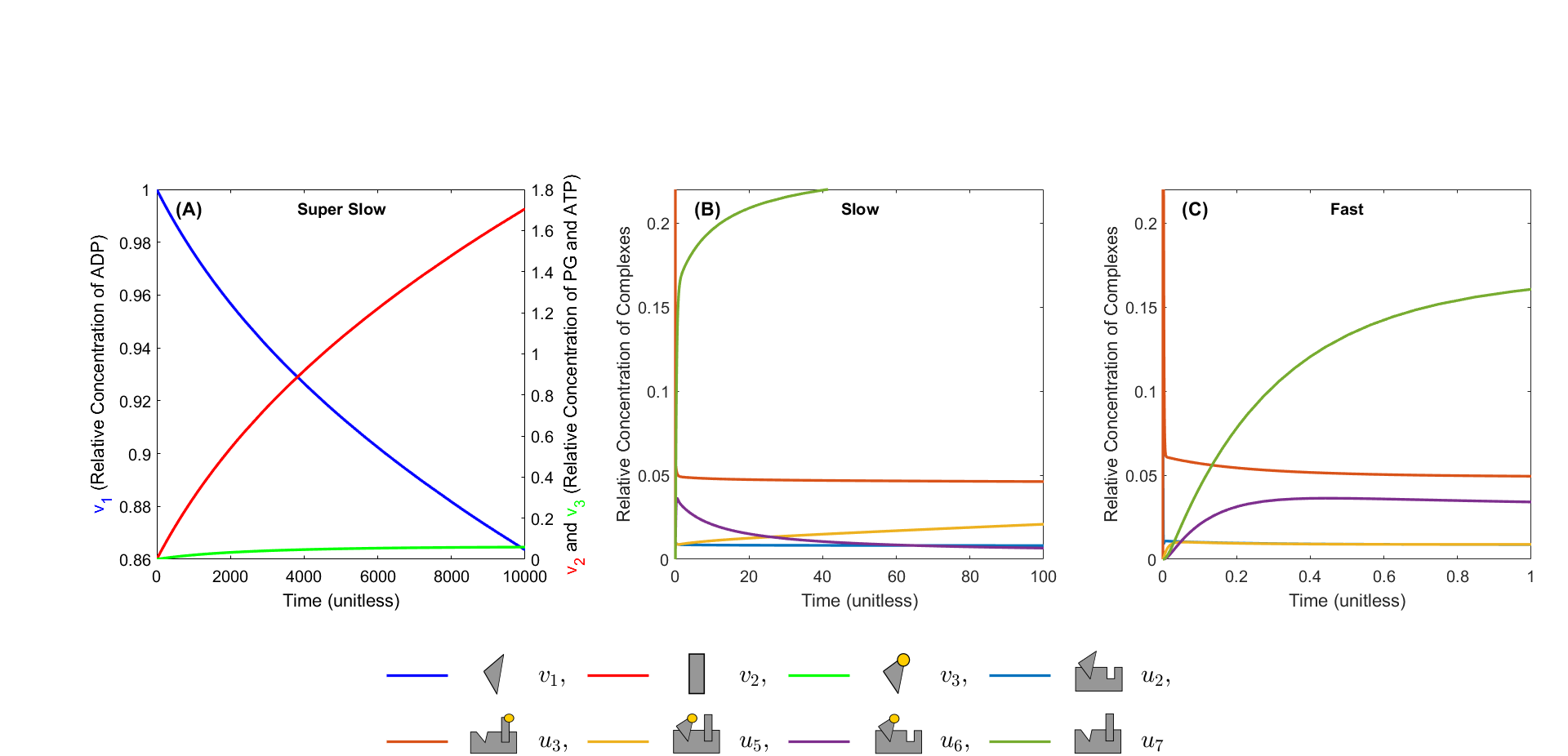}
    \vspace{-15pt}
   \caption{Simulations of the non-dimensionalized model reveal three distinct time scales with no drug input ($z_0 = 0\mu$M). Panel (A) shows super-slow changes to substrate and product concentrations over 10,000 units of time. Panel (B) shows changes in enzyme complexes over 100 time steps. Panel (C) shows the fastest changes over the first unit of time.}
   \label{fig:threescalesnodose}
\end{figure}